\documentstyle[11pt,aaspp4]{article}
\begin{document}
\title{Rotation Curve Measurement using Cross-Correlation \footnote{
Observations reported in this paper were obtained
at the Multiple Mirror Telescope Observatory, a facility operated
jointly by the University of Arizona and the Smithsonian Institution.} } 
\author{Elizabeth J. Barton\altaffilmark{2}, Sheila J. Kannappan, Michael J. Kurtz, Margaret J. Geller}
\affil{Harvard-Smithsonian Center for Astrophysics}
\authoraddr{ Mail Stop 19, 60 Garden St. Cambridge, MA 02138,
        (email: Betsy.Barton@cfa.harvard.edu, skannappan@cfa.harvard.edu, mkurtz@cfa.harvard.edu, mgeller@cfa.harvard.edu)}
\altaffiltext{2}{present address: National Research Council of Canada, Herzberg Institute of Astrophysics, Dominion Astrophysical Observatory, 5071 W. Saanich Road, RR5,
Victoria, BC, Canada V8X 4M6}

\begin{abstract}
Longslit spectroscopy is entering an era of increased spatial
and spectral resolution and increased sample size.  Improved
instruments reveal complex velocity structure that cannot be
described with a one-dimensional rotation curve, yet samples
are too numerous to examine each galaxy in detail.  Therefore,
one goal of rotation curve measurement techniques is to flag cases
in which the kinematic structure of the galaxy is more complex
than a single-valued curve.

We examine cross-correlation as a technique that is easily 
automated and works for low signal-to-noise spectra.  We 
show that the technique yields 
well-defined errors which increase when the simple spectral
model (template) is a poor match to the data, flagging those cases
for later inspection.

We compare the technique to the more traditional, parametric
technique of simultaneous emission line fitting.
When the line profile at a single slit position is
non-Gaussian, the techniques disagree. For our model spectra with two 
well-separated velocity components, assigned velocities from the two 
techniques differ by up to $\sim$52\,\% of
the velocity separation of the model components.  However,
careful use of the error statistics for either technique 
allows one to flag these non-Gaussian spectra. 

\end{abstract}


\section{Introduction}

Spatially resolved optical spectroscopy became a tool for 
studying the dynamics of external galaxies when Pease (1918) 
observed rotation in the inner part of M31 (Rubin 1995).  Later, 
Babcock (1939) effectively measured an optical ``rotation curve'' 
of M31 by measuring velocities of individual nebular regions separately.  
Although required exposure times were very long, somewhat 
larger samples of rotation curves
were amassed in the inner portions of galaxies 
(Burbidge \& Burbidge 1975), and finally in the outer regions.
Now, optical rotation curves of nearby galaxies can be measured
with brief exposure times, enabling the construction of very large
samples of rotation curves for statistical purposes 
(e.g. Rubin et al. 1985; Mathewson, Ford \& Buchhorn 1992; 
Courteau 1997).

Improvements in the spatial and velocity resolution of 
optical spectrographs have revealed complex velocity structure 
in both elliptical and spiral galaxies.  The phenomena
include distinct, nuclear kinematic components in spiral
galaxies (e.g. Marziani et al. 1994; Rubin, Kenney \& Young 1997;
Bureau \& Freeman 1999), and in the cores of elliptical galaxies
(e.g. Franx, Illingworth \& Heckman 1989).

The physics behind the complex velocity structure seen in
emission lines is difficult to unravel. 
Although there are numerous techniques for inferring 
the line-of-sight velocity structure of a stellar system from
a non-Gaussian {\it absorption} line profile (e.g. Rix \& White 1992;
van der Marel \& Franx 1993; Merrifield \& Kuijken 1994), the
analogous measurements for emission-line kinematics are largely
unconstrained. Emitting regions have non-uniform internal velocity structure 
which complicates the structure of the
emission lines (e.g. from individual HII regions; Osterbrock 1989).

To date, estimates of the true kinematic structure
of galaxies with complex emission line profiles are largely
qualitative (e.g. Rubin et al. 1997), and most rotation curve 
reductions assign a single velocity at each slit position.
In most longslit, emission-line spectroscopic studies, velocities are
computed with some form of Gaussian line
fitting (e.g. Keel 1996) or line centroiding or peak 
fitting (e.g. Courteau 1997; Rubin et al. 1997), 
of either the brightest emission line or a subset of 
emission lines simultaneously.  Mathewson et al. (1992) and
M\'{a}rquez \& Moles (1996)  use cross-correlation to measure rotation
curves, but without detailed description of their technique. 

In this era of large, high-quality rotation curve samples,
techniques for rotation curve reduction need to be
re-examined.  Here, we evaluate a little-used but
easily-implemented technique for longslit rotation
curve reduction, cross correlation. We give special attention to the
fact that each aperture along the slit may contain emission from
multiple velocities.  The technique (1) yields
a well-defined response to all line
profiles --- the velocity and error have a clear 
physical relationship to the observed line profile, (2) is 
accurate when extracting redshifts
of low signal-to-noise (S/N) spectra, and (3) is easily
automated.  Focusing on (1),
we compare the technique to the fundamentally different,
parametric technique of Gaussian line fitting.

We test the techniques for redshift measurement in a 
controlled manner, using model spectra with varied
properties.  Our intent is to highlight the most relevant
features of the techniques, not to explore parameter space
exhaustively.  In \S~2, we describe two techniques which 
represent fundamentally different
approaches, cross-correlation and Gaussian
fitting.  In \S~3, we test them with single-aperture
Gaussian and two-component line profiles, to illustrate that
when the line profile is non-Gaussian, different techniques
can yield different results, and to track the error behavior.  
In \S~4, we apply the techniques to two-dimensional model 
rotation curves to illustrate the error behavior in
situations which commonly arise in longslit spectroscopy.  
In \S~5 we briefly discuss applying cross-correlation to
longslit spectral data. We conclude in \S~6. 

\section{Exploring Cross-Correlation and Simultaneous 
Emission Line Fitting}

We evaluate an effective technique 
for redshift measurement, cross-correlation (XC), which is
little-used for emission-line measurements.  We compare it
to an alternative technique,
simultaneous emission line fitting (SEMLF; as in Keel 1996).  
The techniques represent the two fundamentally different approaches to
these measurements. 
SEMLF is parametric, involving modeling the 
emission lines and fitting for the parameters.  
XC is non-parametric
in the sense that we do not fit for parameters, 
although the technique does involve assumptions ---
we use a model spectrum for a template in the computation. 
Our comparison of the techniques shows that when a 
single-velocity measurement is ill-defined, as when
components at two separate velocities contribute to the
spectrum, different approaches yield different results.
Thus, these cases must be flagged and dealt with separately
for a proper characterization of the velocity structure of
each galaxy.

Both techniques can be reliably automated, although SEMLF requires
some fine-tuning, as does XC in the low S/N case.
XC is easily implemented 
with {\bf xcsao} (see Kurtz \& Mink 1998), within the IRAF 
(Tody 1986; 1993) environment.  
We use a version of SEMLF based on that of 
Kannappan et al. (1999) implemented with IDL 
(Landsman 1995).  Both techniques
apply a specific model to the data --- they assume a pre-determined
line profile which is usually a Gaussian.  
However, longslit observations are
designed to detect spatially separated components {\it at different
velocities}. These components may broaden line profiles,
or may turn them into double-peaked profiles. When the Gaussian
model is not a good representation of the data, 
the techniques will produce different 
results.  Below, we explore results for Gaussian and 
double-Gaussian profiles. 

\begin{figure}
\centerline{\epsfysize=7in%
\epsffile{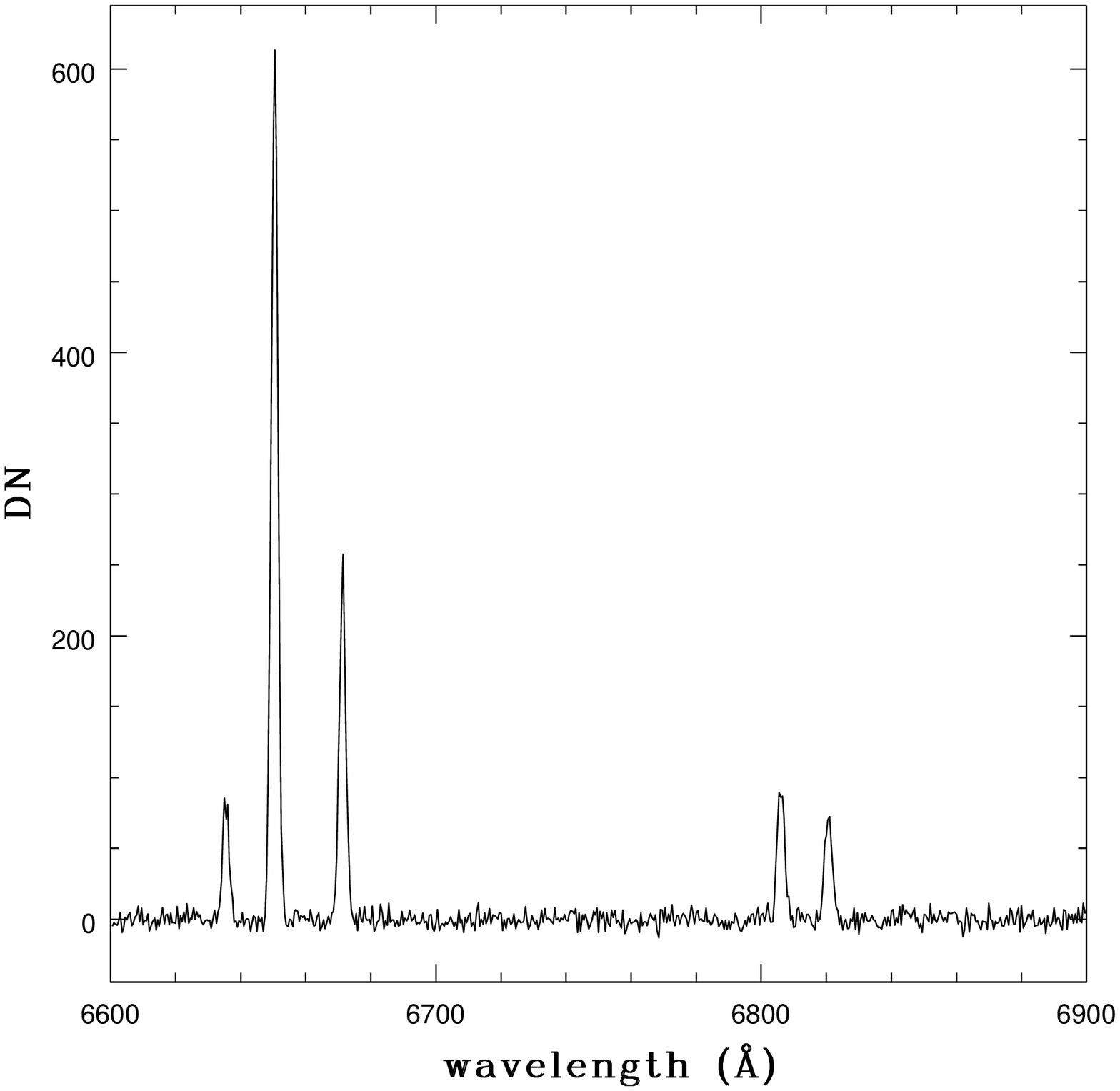}}
\caption{The basic model spectrum, shifted to $4000 {\rm\ km\ s}^{-1}$.}
\label{fg:a1}
\end{figure}

We use artificial emission-line spectra with noise to 
explore and compare the XC and SEMLF
techniques.  We construct the spectra with the {\bf linespec} task in
RVSAO, and add noise with {\bf mknoise} in NOAO.ARTDATA. 
Figure~\ref{fg:a1} shows the basic spectrum, a set of 5 Gaussian
profiles centered at the major emission lines, H$\alpha$, [SII] and
[NII], redshifted to 4000.0 km~s$^{-1}$.  We choose linewidths typical
of spectra taken with a 1200 g/mm grating and a narrow 
(1.0$^{\prime\prime}$) slit --- H$\alpha$ has a full width
at half maximum (FWHM) of
2.3~\AA\ $=$~104~km~s$^{-1}$.  The other lines
satisfy 2.26 ~$\leq {\rm FWHM} \leq 2.50$~\AA.  To
add the noise, we assume a gain of 1.5 e$^{-}$/DN and a read noise of
7.0 e$^{-}$. Although the continuum S/N ratio is not defined in
these model spectra, the effective signal-to-noise ratio of the
H$\alpha$ line is very large in many of them: 
the H$\alpha$ signal in the spectrum in Fig.~1
is $\sim$2300~e$^{-}$ over 5 pixels, so the S/N ratio in H$\alpha$ is 
$\sim$45.  Although we vary the S/N ratio in the model spectra
in the examples of \S~3.1, most of the other test spectra have
comparably large S/N ratios.  The discrete pixel sampling scale is
$\sim$22~km~s$^{-1}$.

\subsection{The Cross-Correlation Technique}

Kurtz \& Mink (1998) describe cross-correlation exhaustively for
application to redshift surveys, using the task {\bf xcsao} in the
RVSAO package, within the IRAF environment (Tody 1986, 1993). They
include template construction and error analysis in their discussion. 
We apply the XC technique using {\bf xcsao} in RVSAO, according to the
procedure that would be used for actual data (excluding bias
subtraction, flat-fielding, wavelength calibration,
and cosmic ray removal).
We construct the template based on the median widths and
relative heights of the ``emission lines'' in a large ensemble of
model spectra, just as we would for actual data (see \S~4).
Because the high-resolution models are undersampled relative
to the optimal sampling rate for {\bf xcsao}, 
we must adjust the rebinning in {\bf xcsao}, by using both
linear and spline3 interpolation and choosing the best result
based on the $r$ statistic.

\begin{figure}
\centerline{\epsfysize=7in%
\epsffile{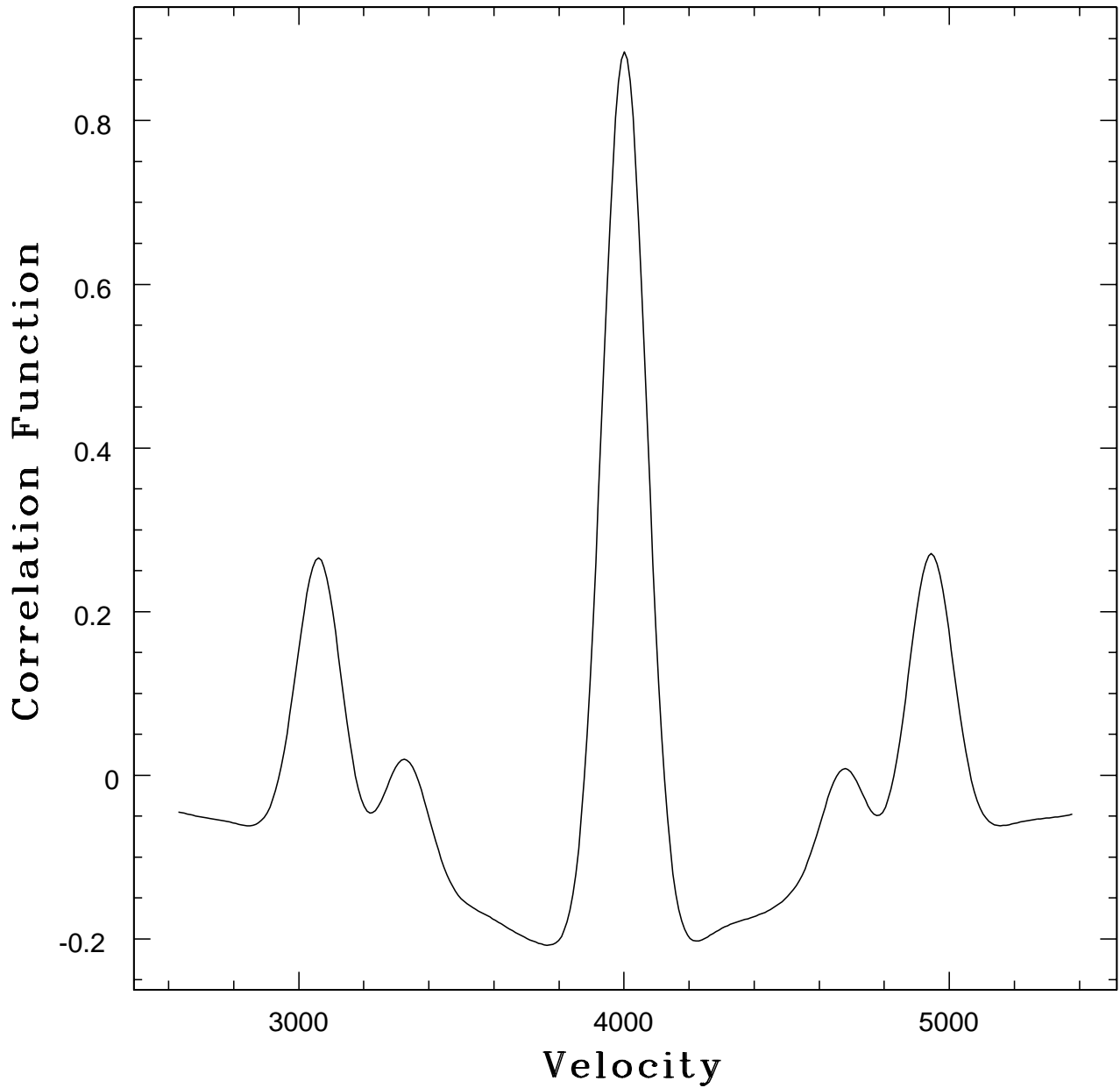}}
\caption{The correlation peak of the large signal-to-noise spectrum
in Fig.~1, fit by {\bf xcsao} to determine the XC velocity.}
\label{fg:a2}
\end{figure}

Fig.~\ref{fg:a2} shows a sample peak in the
correlation function, which {\bf xcsao} fits to find the velocity
for the spectrum in Fig.~\ref{fg:a1}.  
Because the sample spectrum has a very large S/N ratio and
a nearly-perfect Gaussian profile,
the peak is well-defined and the redshift,
$4001.1 \pm 0.4 {\rm\ km\ s}^{-1}$, is close to the input model
value of $4000 {\rm\ km\ s}^{-1}$. 

Cross-correlation errors include the effects of spectrum/template
mismatch, and therefore have a clear relationship to the spectral
profile.
The internal error estimator of Kurtz \& Mink (1998) follows from the
discussion in Tonry \& Davis (1979, TD hereafter), with the additional
assumption that the noise is sinusoidal. {\bf xcsao} computes an
error of $\frac{3}{8}\frac{\omega}{1+r}$, where $\omega$ is the FWHM 
of the correlation peak and $r$, defined by TD,
is a measure of the noise based on the antisymmetric part of the
correlation function.
TD assume that the (symmetric) template, 
convolved with a simple symmetric function, is a
noise-free spectral match to the object --- that all
spectrum/template mismatch is the result of noise.  
Under these assumptions, the antisymmetric
part of the correlation function yields a measure of the height of the
average noise peak --- a galaxy with symmetric line profiles,
cross-correlated with a template with symmetric line profiles, 
has noise and no signal in the antisymmetric part of its correlation
function. Thus, the error that {\bf xcsao} computes is the shift 
in the correlation peak center that would result from a 
spurious noise peak, where
$\frac{3}{8}\omega$ is the average distance to the nearest noise peak
and, with the assumptions of TD, $r$ is an estimate of the 
height of the true correlation peak
divided by of the height of the average noise peak. 

If $r$ is small, there are spurious
peaks in the correlation function comparable to or exceeding
the highest peak.  {\bf xcsao} is likely to fit one of these 
spurious peaks instead of the true peak; 
then the computed velocity is arbitrarily
far from the true velocity, much farther than the formal error
indicates.  Thus, $r$ is a measure of the reliability of the
redshift.
Kurtz \& Mink (1998) require $r > 3.0$ for automatic acceptance
of the extracted redshift.  
However, redshifts with $r < 3.0$ can be used
in longslit reduction because there are data in neighboring
apertures.  If {\bf xcsao} fits the wrong peak, the redshift
will appear discrepant from neighboring apertures and the reduction can
be checked manually, or rejected. 

If the template has a single velocity component, the assumptions underlying
the error estimator are incorrect for spectra with multiple components
at different velocities. In this multiple-component case,
spectrum/template mismatch results from noise {\it and} from additional
peaks in the spectrum at different velocities.  The mismatch generally
appears in the antisymmetric part of the correlation function, and $r$
{\it measures this template mismatch}. Thus $r$ becomes a measure of how well
the template fits the spectrum.  Low $r$ values indicate a poor fit,
and result in large error.  $\omega$ also reflects additional nearby
velocity components; an additional nearby component results in a wider
correlation peak, a larger $\omega$ and a larger error.  Therefore,
additional velocity components enlarge XC errors significantly.  Below,
we explore this effect using model spectra. 

Kurtz \& Mink (1998) find that $\frac{3}{8} \frac{\omega}{1+r}$
systemically over or underestimates nuclear 
redshift errors by $\sim$20\,\%. For
large redshift surveys consisting of nuclear spectra reduced uniformly
with a single template, $\omega$ remains roughly constant. In that case,
error calibration can be applied to eliminate the discrepancy. 
Kurtz \& Mink (1998) solve for a template-dependent constant, $k$,
where the true error is then $\frac{k}{1+r}$. 

However, in longslit spectroscopy, where multiple-velocity spectra
abound, $\omega$ is not constant. Throughout this paper, we use
$\frac{3}{8} \frac{\omega}{1+r}$ to estimate XC errors. We recommend
computing errors proportional to $\frac{\omega}{1+r}$ for all longslit
reduction implemented with XC, because $\omega$ reflects the presence
of multiple velocity components and wide velocity components. 

\subsection{The SEMLF Technique}

The major steps involved in SEMLF reduction are similar to the steps
required to implement cross-correlation, except that we measure
velocities by fitting single-Gaussian functions to the major emission
lines simultaneously, when the lines are detected at $\geq 3 \sigma$. 
All spectra reduced with the SEMLF program were first transformed to
log($\lambda$) space using the task {\bf transform} in
the LONGSLIT package of IRAF, with an artificial calibration
lamp image.  
In our implementation of SEMLF, the relative wavelengths are fixed
and the linewidths are constrained to be the same value for each line;
the overall linewidth and each individual peak height may vary.
We derive the formal model-dependent errors from $\chi^2$-minimization
fitting using the Gaussian 
model (Press et al. 1992).  

When the minimal $\chi^2 \neq N \pm \sqrt{N}$, where 
$N$ is the number of degrees of freedom in the fit, the 
formal errors cannot be justified rigorously. 
Press et al. (1992) emphasize that
these errors are unsuitable when a model is an incorrect
representation of the data.  Finding the true errors
requires other methods (e.g. Monte Carlo simulation) which are
usually computationally expensive.  
Authors generally compute errors that account for photon statistics; some
error calculations are independent of profile shape (Courteau 1997 uses
the weighted mean) 
and/or include line widths (Keel 1996).  However, these errors do not
include mismatch between the data and a basic model that in every case
assumes a single, well-defined velocity.  Here, we attempt to partially 
account for this mismatch, taking an approach guided by the analogy between 
XC errors and the error derived 
from weighting by the reduced $\chi^2$, because $\chi^2$ is an estimator 
of the suitability of the Gaussian model and reflects irregular 
line profiles. Although the procedure is not rigorously justified, 
we consider weighted errors as a computationally convenient way of
achieving an estimate of the proper error behavior; below, we multiply the error
by the reduced $\hat{\chi}^2 = \chi^2/N$. We demonstrate that the 
weighted errors display the expected error behavior and are 
similar to the XC errors.  This similarity arises because both the
XC $\frac{1}{1+r}$ statistic and the SEMLF $\hat{\chi}^2$ are
based on the cross product of the object and model spectra
divided by an estimator of the variance.  The formal SEMLF error
performs the same function as the width factor ($\frac{3}{8}\omega$)
in {\bf xcsao}; it provides a scale by which to multiply the goodness of
fit measure ($\hat{\chi}^2$ for SEMLF, $\frac{1}{1+r}$ for XC) in
order to estimate the uncertainty of the measurement.

The $3 \sigma$ cutoff of SEMLF is analogous to imposing a lower limit
on $r$ in the XC technique.  The limit is somewhat arbitrary and could
be modified.  
In general, we do not modify the limit in our tests, so these tests
do not directly compare the effectiveness of 
the two techniques on very low S/N spectra.  

We apply the SEMLF technique to the spectrum of Fig.~\ref{fg:a1},
with a redshift of 4000~km~s$^{-1}$.
The resulting fit, with a velocity and
weighted error of $4000.9 \pm 0.5 {\rm\ km\ s}^{-1}$, is
nearly indistinguishable from the spectrum.  Thus,
both SEMLF and XC find the correct velocity of the line profiles 
when they are Gaussian (with large S/N ratios).
SEMLF measures the proper error in the Gaussian case;
thus SEMLF, used with calibration spectra consisting of Gaussian
line profiles, provides one method of calibrating the XC error
when necessary.

\section{Behavior of the Two Techniques for One-Dimensional Spectra}

In the following sections, we explore the behavior of XC and SEMLF. 
We use one-dimensional spectra 
with varying line profiles. We fix the emission line ratios 
to standard HII region values similar to the XC template.
Later, we consider two-dimensional spectra which
mimic different features of galaxy rotation curves, including varying
emission-line ratios.  All models
match the resolution of the data described in \S~3.
However, most of the model spectra have larger SNR's, and
therefore much larger $r$ values, than typical rotation curve data.

\subsection{Gaussian Line Profiles}

\begin{figure}
\plottwo{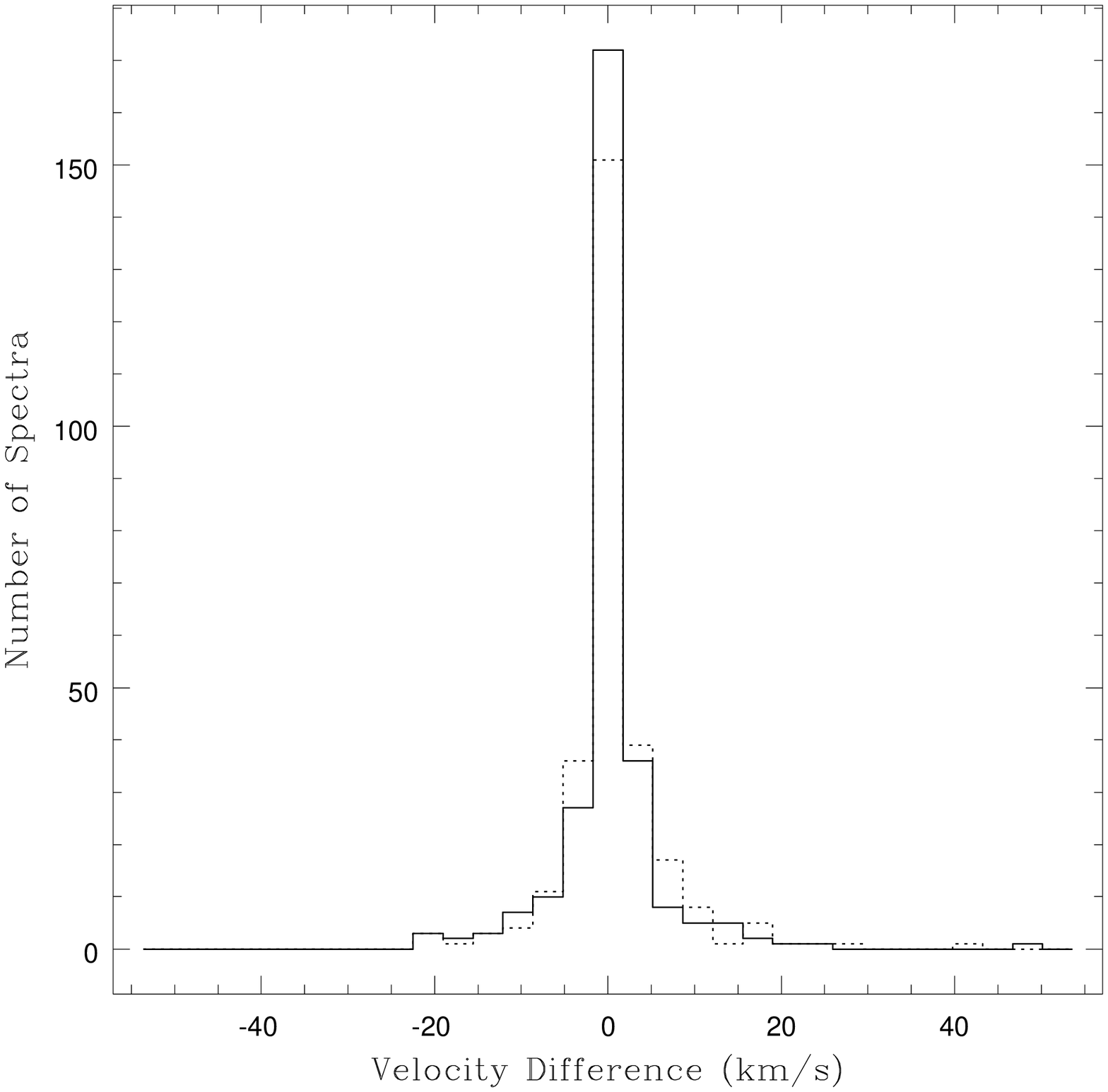}{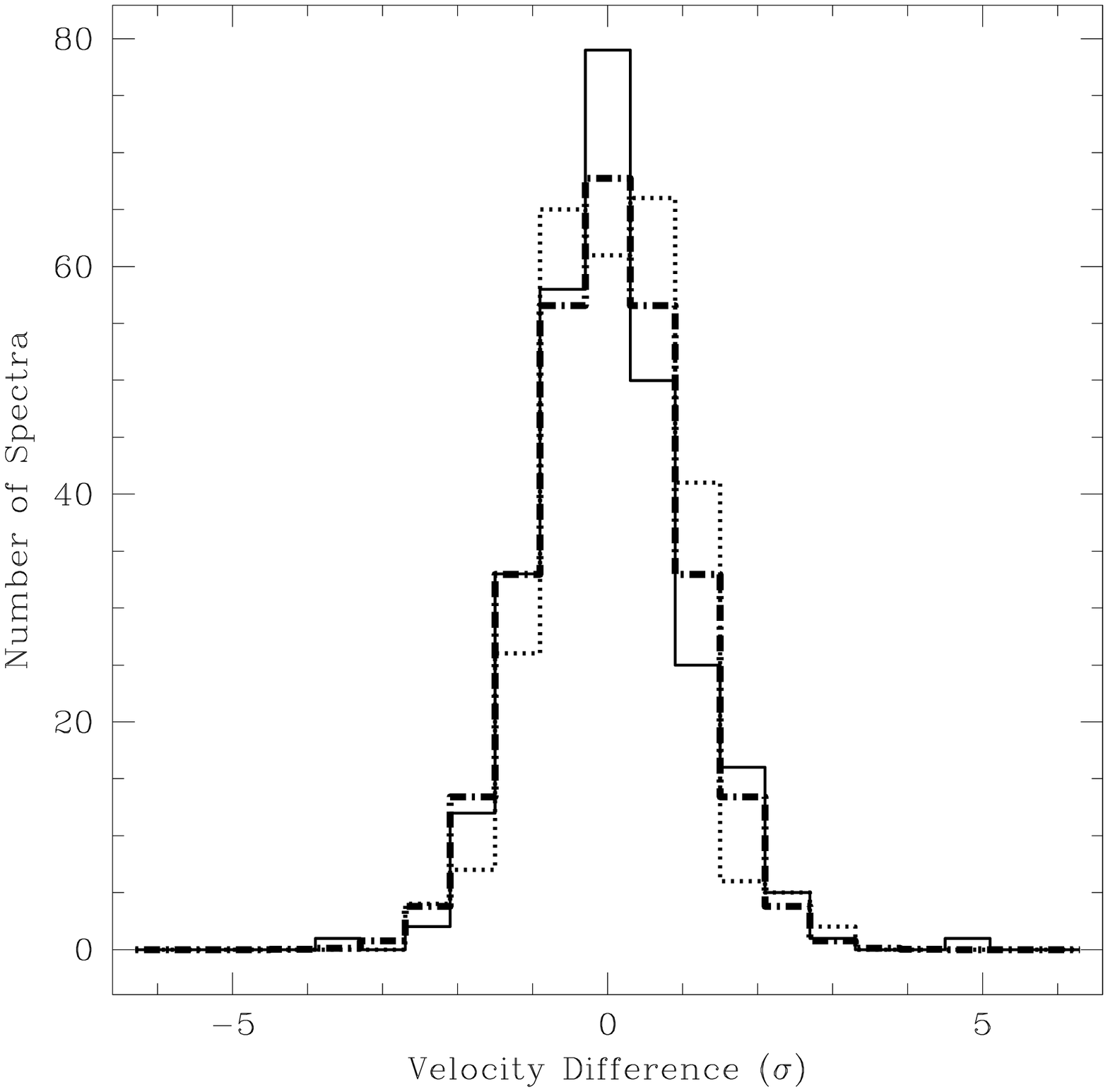}
\caption{283 Gaussian spectra at different noise levels: (a) histogram
of the difference between the true velocity and the XC velocity
(solid line), and the true velocity and the SEMLF velocity
(dotted line) in km/s and, (b) histogram
of the difference between the true velocity and the XC velocity
divided by the XC error (solid line), or the true velocity and the
SEMLF velocity divided by the $\hat{\chi}^2$-weighted SEMLF error 
(dotted line).  The thick
dot-dashed line is a Gaussian distribution with $\sigma=1$.}
\label{fg:a3}
\end{figure}

We apply XC and SEMLF to a set of 400 single-Gaussian spectra identical
to Fig.~\ref{fg:a1}, except that H$\alpha$ S/N ratios vary from
$\sim$54 to $< 1$.
XC finds a velocity with $r \geq 1.5$ for 297 of the
spectra; SEMLF finds a redshift for 285 using the $3\sigma$ cutoff; 
we use the 283 spectra with results from both techniques for our analysis. 
Fig.~\ref{fg:a3}a shows a histogram of the difference between the XC
velocity and the input model velocity (solid line), superimposed on
the difference between the SEMLF velocity and the model velocity
(dashed line) for these 283 spectra.  The two distributions are
similar.  Fig.~\ref{fg:a3}b shows the same quantity divided by the
error for each velocity measurement; the dotted line reflects 
$\hat{\chi}^2$-weighted
SEMLF errors; the graph is very similar with unweighted errors.  
If errors were Gaussian and were
properly estimated, the histograms would be Gaussian with $\sigma=1$
(shown as the thick dot-dashed line).  The results of the two techniques
differ little in the case of simple, Gaussian line profiles. 

These results appear to conflict with those of
Kurtz \& Mink (1998), who find that cross-correlation 
is more sensitive than emission
line fitting --- {\bf xcsao} found reliable velocities for many test
spectra which could not be fit with {\bf emsao}, the automatic emission
line fitting routine in RVSAO.  However, the line 
fitting algorithm in {\bf emsao} differs substantially from the SEMLF 
technique of this paper --- we apply constraints to the model we fit,
namely the relative wavelengths and widths of the emission
lines, thus enabling the fitting of spectra with lower S/N. 
Therefore, velocities for low S/N spectra should be
computed with techniques that assume these or similar constraints.

\subsection{Double-Gaussian Line Profiles}
A single, unresolved velocity component with a symmetric line profile
usually dominates a wide-aperture nuclear spectrum of a galaxy; thus 
techniques that assign a single, systemic velocity based on the center
of a galaxy are appropriate.  However, longslit observations are
designed to detect spatially separated components {\it at different
velocities}.  If components at distinct velocities within the same
seeing disk are separated by a small velocity difference and are
smoothly varying, they may only broaden the line profile. One example
is the unresolved but rapidly rotating inner region of a galaxy
with a circumnuclear gas disk.  However, if components are separated by a
large velocity they produce a multi-peaked line profile.  Examples
include the transition region between a rapidly rotating circumnuclear
gas disk or a bar and the outer regions of a galactic disk (Rubin 
et al. 1997).  In this case, the spectrum shows two distinct velocity
components. 
  
When line profiles are non-Gaussian, XC may yield a
different velocity than either SEMLF or centroiding techniques. 
Because the goal of
these reduction techniques is to assign a single velocity to a
spectrum, none of the techniques used for rotation curve reduction,
including cross-correlation, completely model the system in the
multiple-component case.  At best, an automated technique can 
reflect the presence of more than one
component in the computed error statistics. 

\begin{figure}
\plottwo{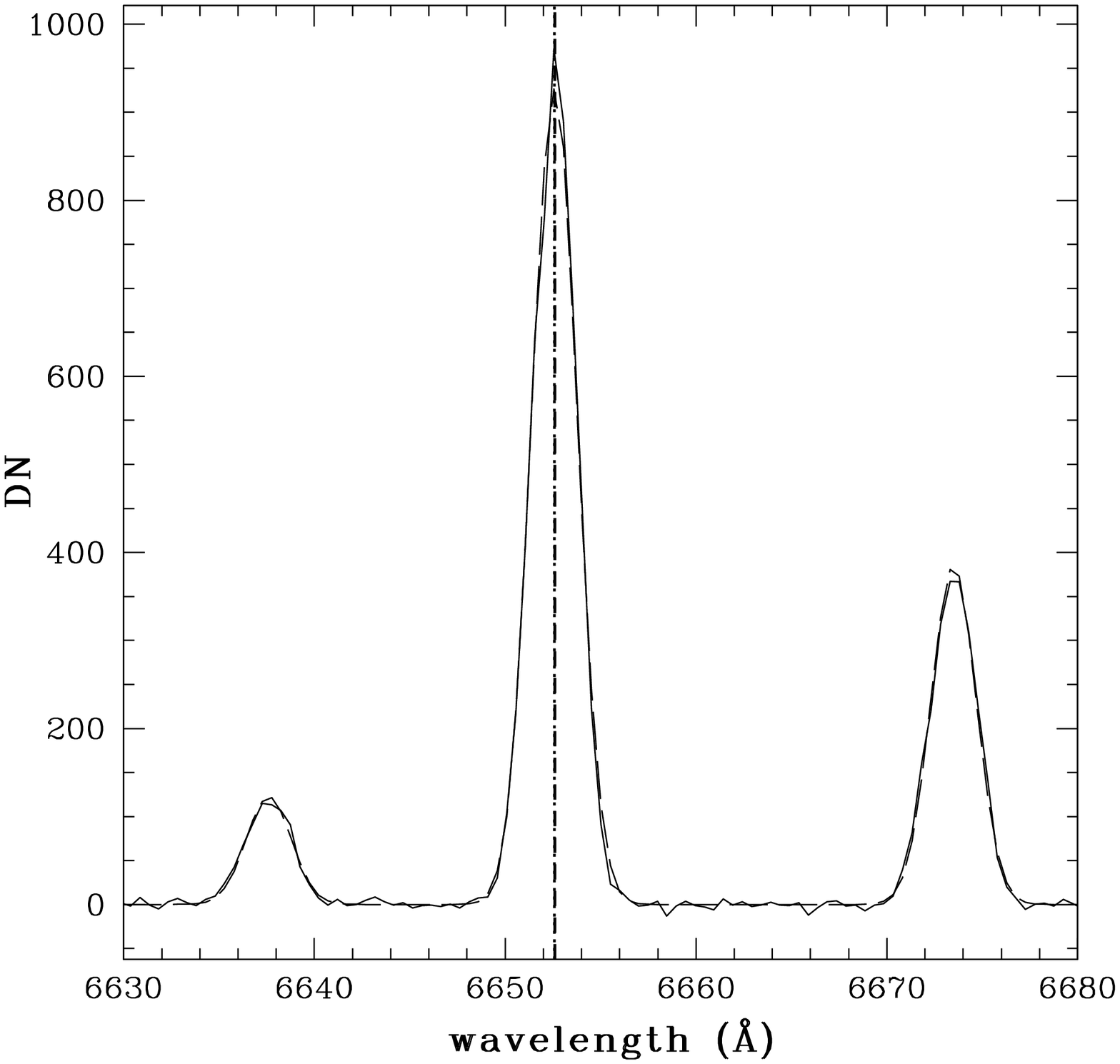}{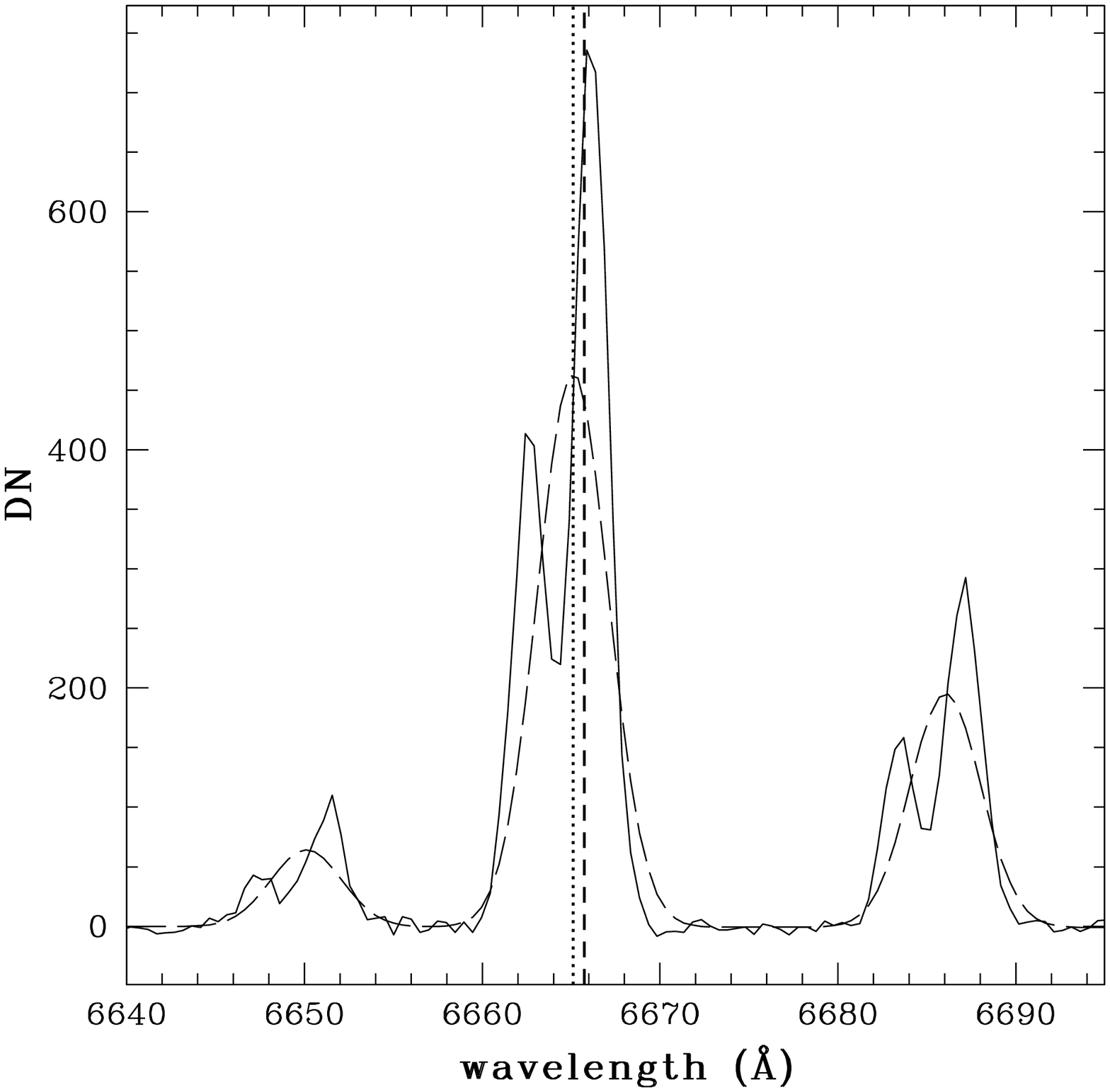}
\caption{Two-component spectra with small and large component
velocity separations, $\Delta V$. The component flux ratio is 2:1.
Each figure shows the
spectrum (solid line) with SEMLF fit (dot-dashed line) and, 
XC (vertical dashed line) and SEMLF (vertical dotted line) results.
(a) $\Delta V = 
60 {\rm\ km\ s}^{-1} = 0.58\ {\rm FWHM}_{\rm H\alpha}$, where
${\rm FWHM}_{\rm H\alpha}$ is the FWHM of the
H$\alpha$ line. The brighter component is at 4120~km~s$^{-1}$.
The XC and SEMLF results overlap; they are
$4101.2 \pm 0.3$ and $4100.8 \pm 0.3 (\pm 0.5)$ km/s, respectively,
where the second SEMLF error is weighted by $\hat{\chi}^2$.
(b) $\Delta V = 160 {\rm\ km\ s}^{-1} = 1.5 {\rm\ FWHM}_{\rm H\alpha}$.
The brighter component is at 4720~km~s$^{-1}$; the XC and SEMLF
results are $4701.3 \pm 4.6$ and $4672.8 \pm 0.7 (\pm 10.9)$ km/s.}
\label{fg:a4}
\end{figure}

\begin{figure}
\plottwo{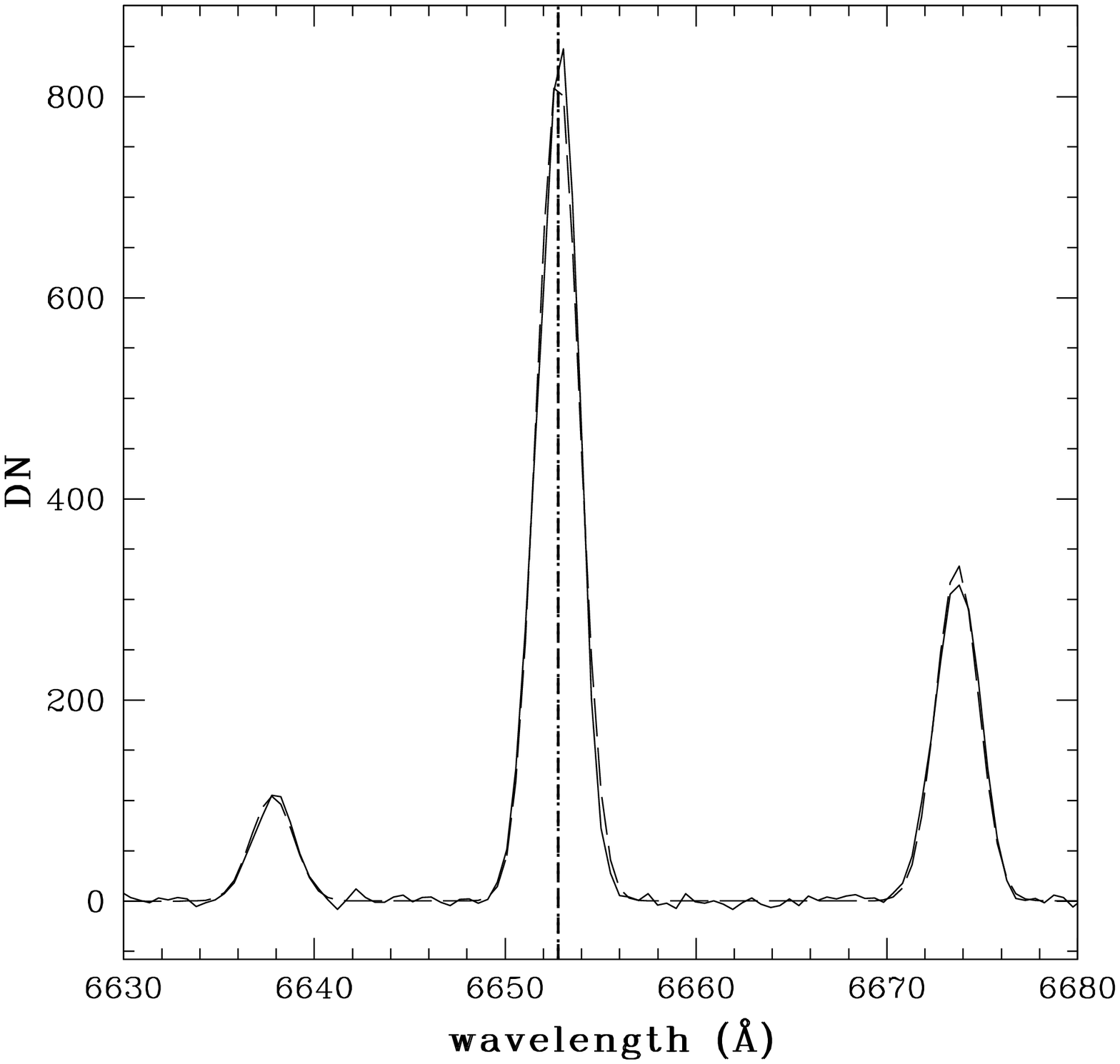}{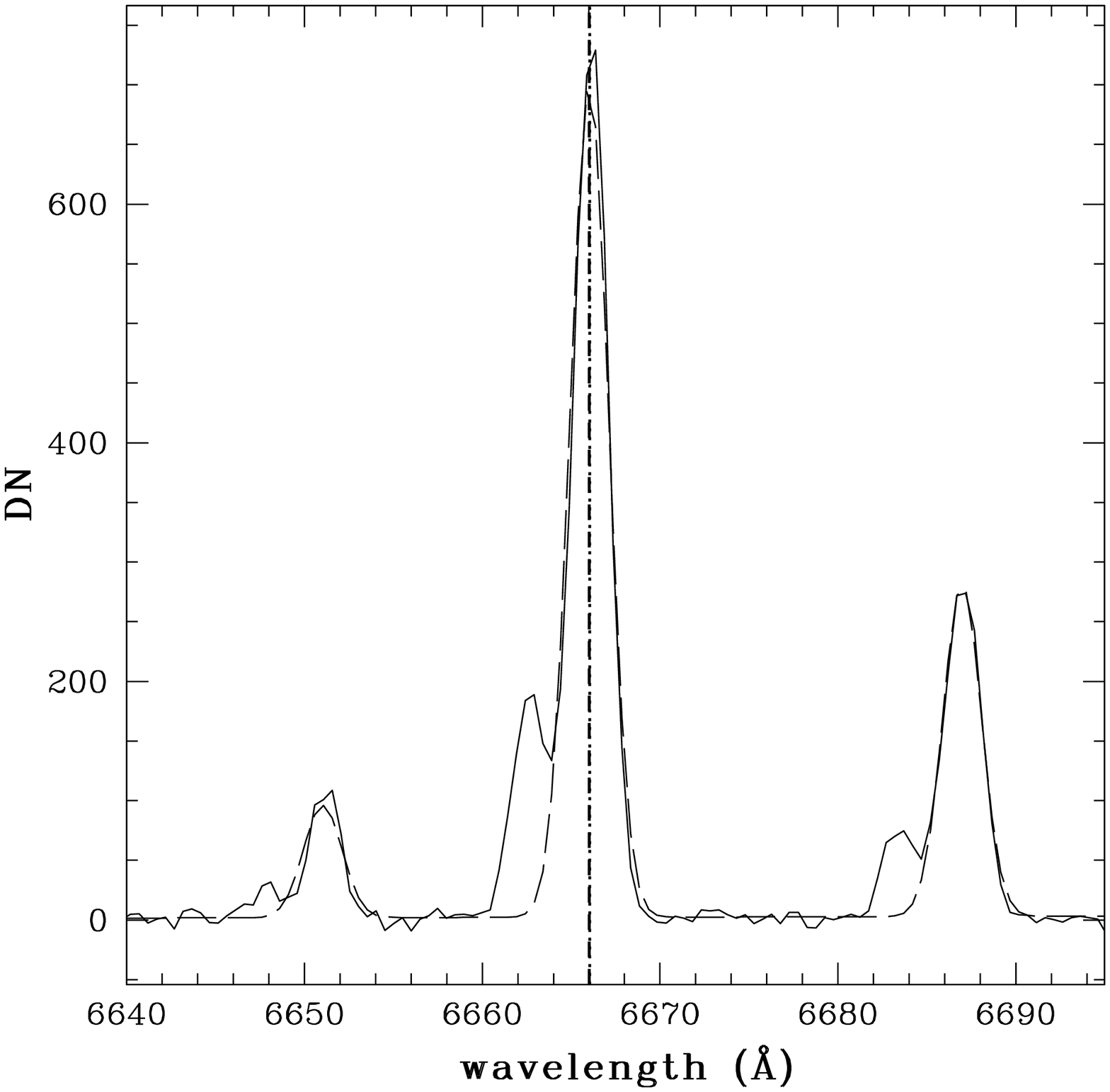}
\caption{Two-component spectra. The component flux ratio is 4:1;
the figure shows the spectrum (solid line) with the SEMLF 
fit (dot-dashed line) and, 
XC (vertical dashed line) and SEMLF (vertical dotted line) results.
The component velocity separations are (a) $\Delta V = 
60 {\rm\ km\ s}^{-1}= 0.58\ {\rm FWHM}_{\rm H\alpha}$, where
${\rm FWHM}_{\rm H\alpha}$ is the standard deviation of the
H$\alpha$ line. The brighter component is at 4120~km~s$^{-1}$.
The XC and SEMLF results overlap; they are
$4109.6 \pm 0.3$ and $4109.3 \pm 0.3 (\pm 0.5)$ km/s respectively,
where the second SEMLF error is weighted by $\hat{\chi}^2$, 
and (b) $\Delta V = 160 {\rm\ km\ s}^{-1} = 1.5
{\rm FWHM}_{\rm H\alpha}$.  The brighter component is at
4720~km~s$^{-1}$; the XC and SEMLF results are
$4714.0 \pm 1.8$ and $4715.0 \pm 0.4 (\pm 3.0)$ km/s.}
\label{fg:a5}
\end{figure}

To examine the behavior of the reduction methods in the two-component case,
we construct spectra equivalent to Fig.~\ref{fg:a1}, except that they
have two distinct Gaussian velocity components. 
Figs.~\ref{fg:a4}~--~\ref{fg:a5} show spectra where the intensity
ratios of the two velocity components are
2:1 (Fig.~\ref{fg:a4}) or 4:1
(Fig.~\ref{fg:a5}).
In each case, the left part of the figure shows components
separated by $\Delta V = 60 {\rm\ km\ s}^{-1}$, or 0.58 of the
FWHM of the H$\alpha$ line
(104~km~s$^{-1}$),
and the right side shows components separated by
$\Delta V = 160 {\rm\ km\ s}^{-1}$, or 1.5 times the
FWHM of H$\alpha$.  The figures 
show the ``spectrum'' around H$\alpha$ (solid line) 
and the SEMLF fit (long-dashed line), along with the output
XC (dashed vertical line) and SEMLF (dotted vertical line) velocities. 
Note that when the velocity resolution differs
from the model spectra we present (the FWHM of H$\alpha$
is 104~km~s$^{-1}$), $\Delta V$ must be scaled
to compare with our results.

In Fig.~\ref{fg:a4}a, the two
separate components are not visible as separate peaks.  SEMLF and XC
yield very similar results.  In Fig.~\ref{fg:a4}b, $\Delta V = 160 {\rm\
km\ s}^{-1}$ --- two distinct peaks are visible.  The SEMLF technique
fits one wide Gaussian to the two components.
In contrast, XC finds a velocity closer to the stronger peak.
The XC error increases by a factor of $\sim$16 over the result in 
Fig.~\ref{fg:a4}a, signaling the presence of the two
distinct components.  Appropriately, $r$ decreases substantially,
from 192 to 14, and and $\omega$ increases from 147 to 182.  The SEMLF 
``formal'' error 
in the $\Delta V = 160 {\rm\ km\ s}^{-1}$ case is
only $\sim$ 2 times larger than in the $\Delta V = 60 {\rm\ km\
s}^{-1}$ case, but the $\hat{\chi}^2$ -- weighted error is $\sim$23
times larger. 

We note a similar trend for the spectral components with a larger
flux ratio (4:1) in  Fig.~\ref{fg:a5}a ($\Delta V = 60
{\rm\ km\ s}^{-1}$) to Fig.~\ref{fg:a5}b ($\Delta V = 160 {\rm\ km\
s}^{-1}$), except that both techniques fit close to the velocity of
the brighter peak ($4720 {\rm\ km\ s}^{-1}$). 
In Fig.~\ref{fg:a5}b, SEMLF primarily fits the brighter peak, and
the secondary peak, which does displace the resulting line
profile, is not strongly reflected in the unweighted error.  On the
other hand, the XC error and the SEMLF weighted error are $\sim$5 and
6 times larger, respectively, than the errors in Fig.~\ref{fg:a5}a,
indicating model mismatch.
Again, $r$ decreases substantially from 164 to 29; $\omega$ stays roughly
the same.

\begin{figure}
\centerline{\epsfysize=7in%
\epsffile{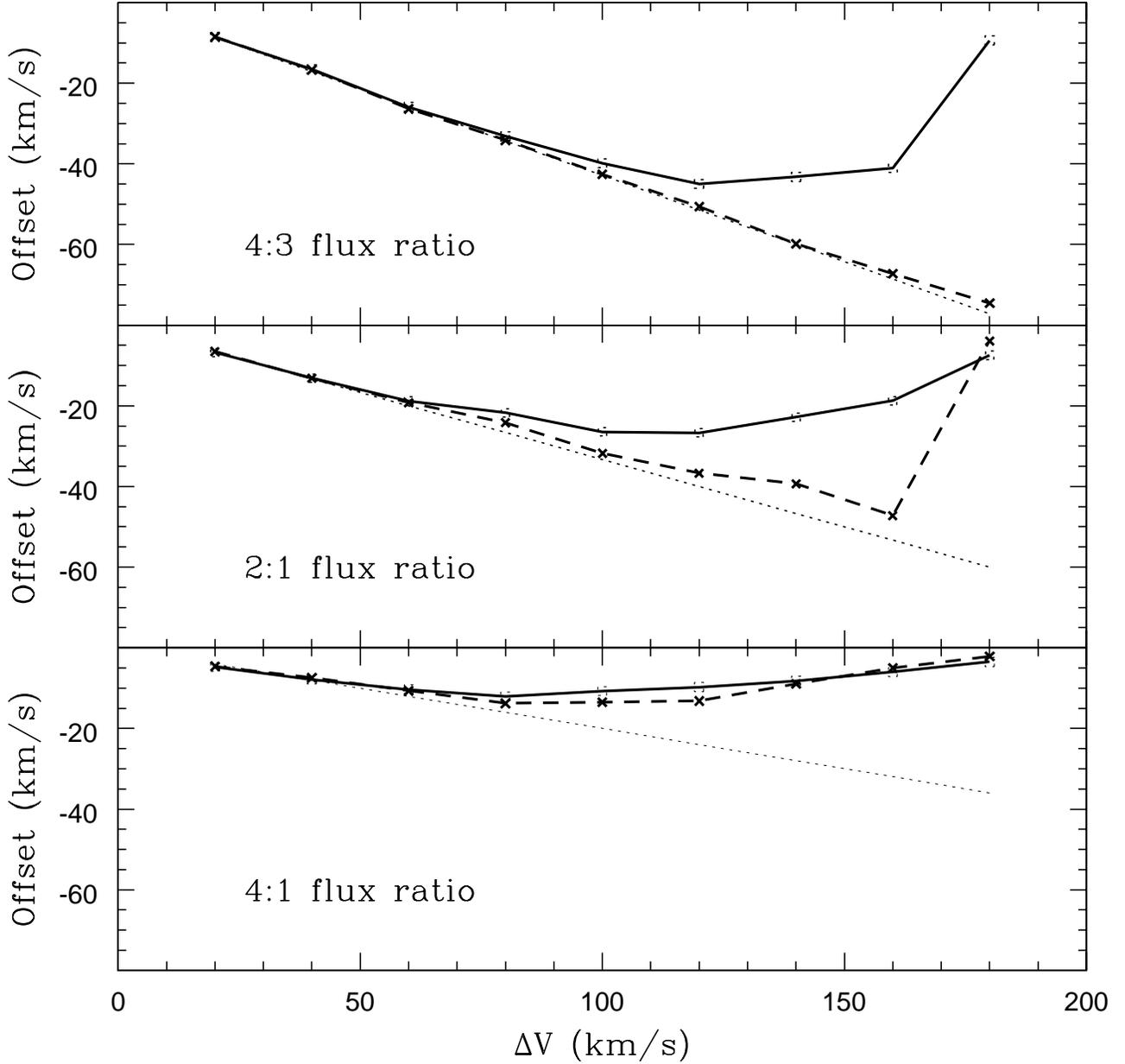}}
\caption{Velocity offsets as a function of velocity component
separation from two-component spectra 
for XC (solid line) and SEMLF (dashed line).  The velocity offset
is equal to the velocity solution
minus the true velocity of brightest component in the model.
The thin dotted line is the flux-weighted mean of the two
components.
The flux ratios are 4:3 (top), 2:1 (middle), and 4:1 (bottom).
For different spectral resolutions, $\Delta V$ must
be scaled by 104~km~s$^{-1}/$FWHM, 
where FWHM is for the the H$\alpha$ line.}
\label{fg:a6}
\end{figure}

\begin{figure}
\centerline{\epsfysize=7in%
\epsffile{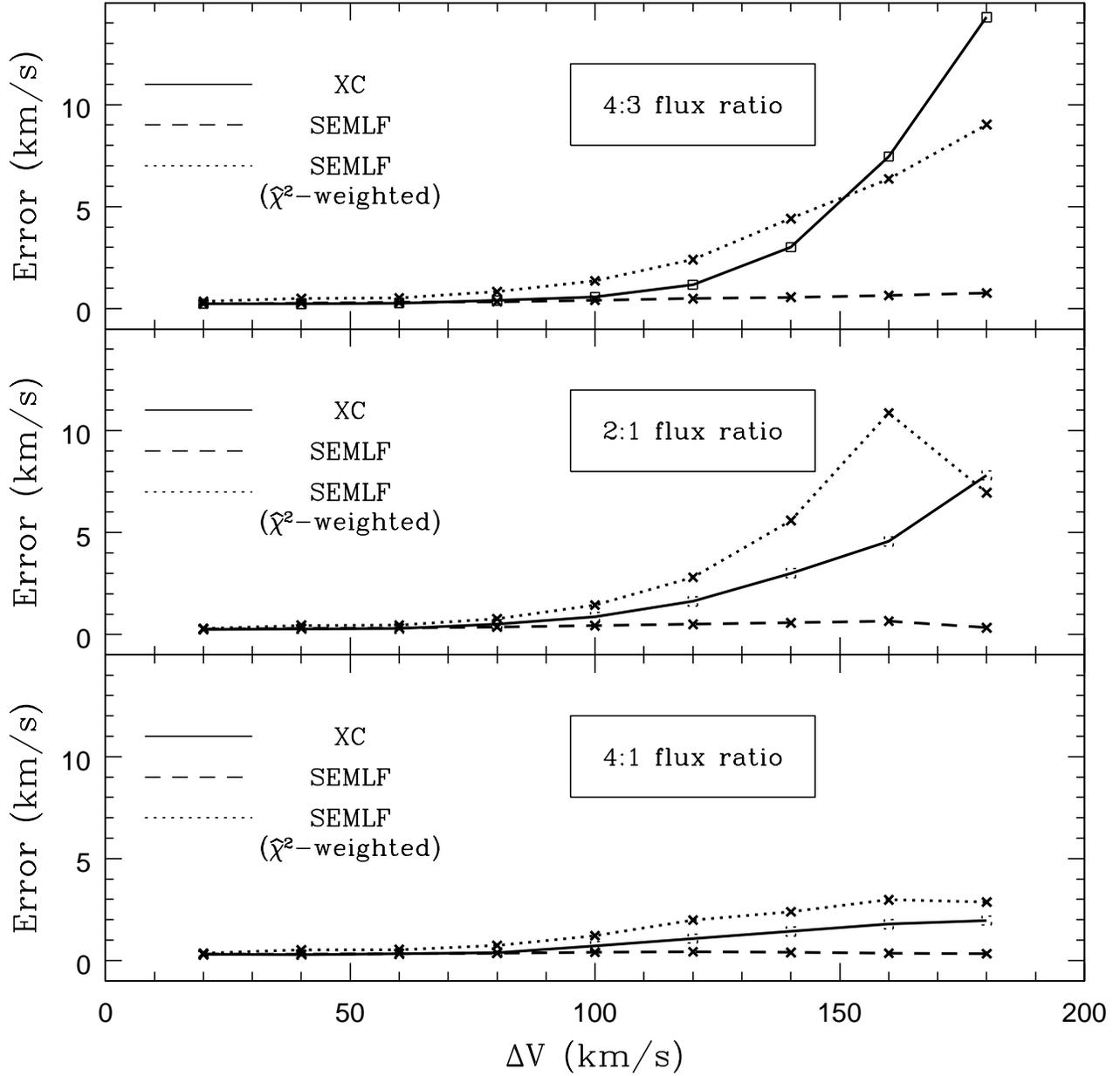}}
\caption{Velocity errors as a function of velocity component
separation for XC (solid line) and SEMLF (dashed line).  As in
Fig.~6, the flux ratios are 4:3 (top), 2:1 (middle), and 4:1 (bottom).}
\label{fg:a7}
\end{figure}

\begin{figure}
\plottwo{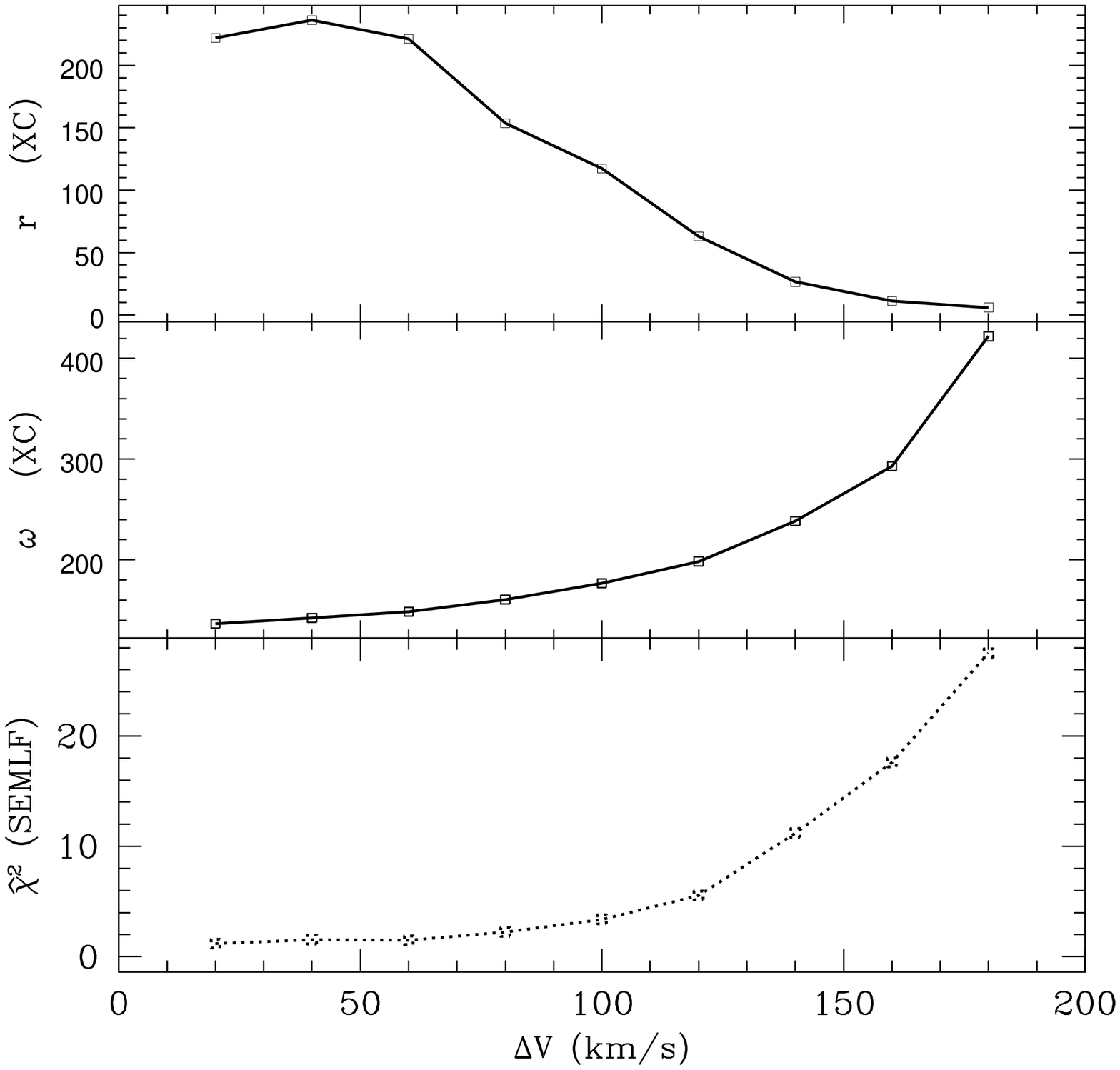}{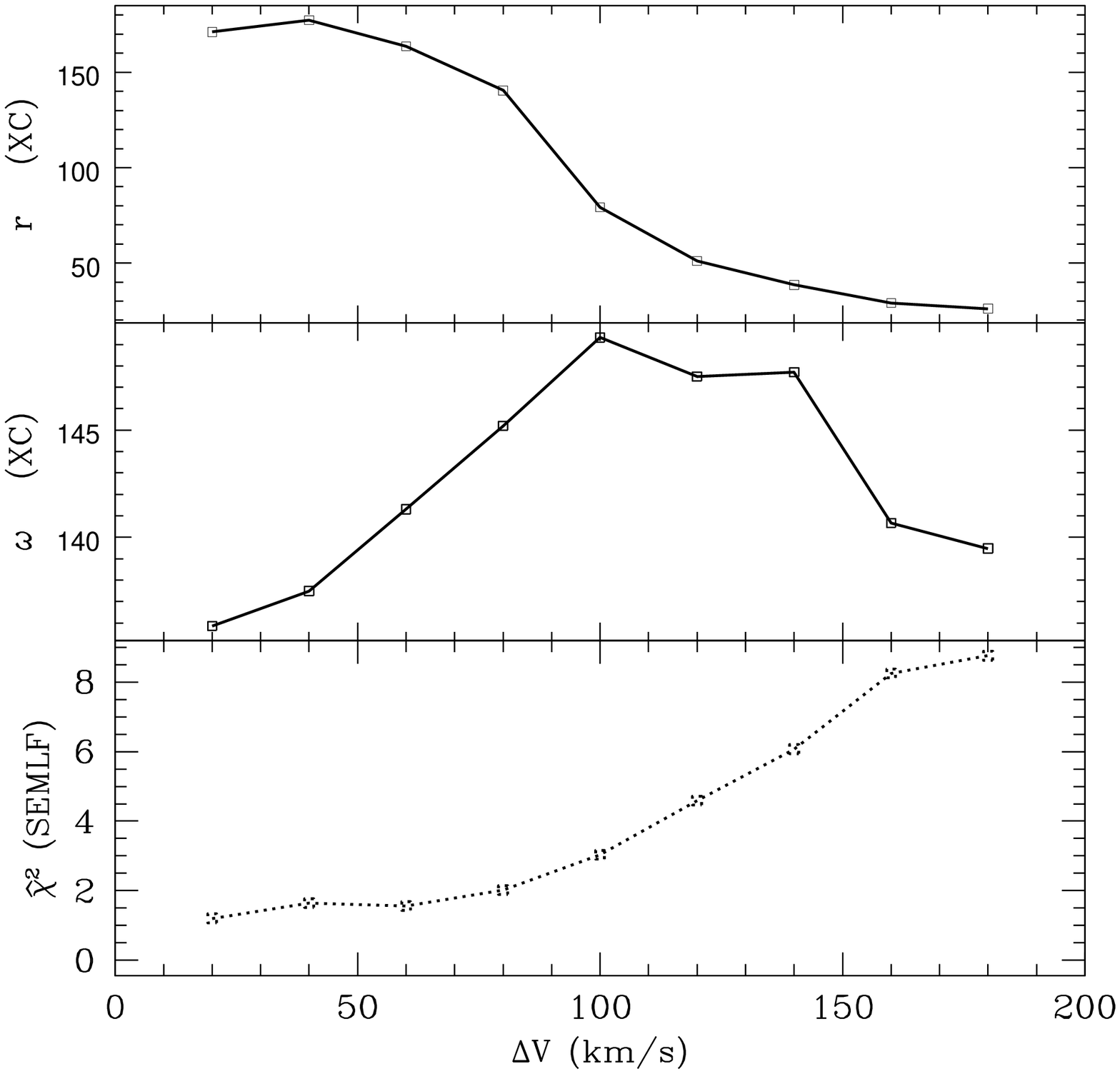}
\caption{Error contributions:
$r$, $\omega$ and $\hat{\chi}^2$ as a function of velocity component
separation, $\Delta V$, for (a) the 4:3 flux ratio and (b) the
4:1 flux ratio. We omit the 2:1 flux ratio case, which looks qualitatively
similar to (b).}
\label{fg:a8}
\end{figure}

Figs.~\ref{fg:a6}~--~\ref{fg:a8} explore these trends for a range of
$\Delta V$.  The model velocity components 
have flux ratios of 4:3, 2:1 or 4:1; the
brighter component is always at the larger velocity.  The model spectra
for each component are identical to the spectra above (e.g.
Fig.~\ref{fg:a5}b provides the points with $\Delta V = 160 {\rm\ km\
s}^{-1}$ and flux ratio 4:1 for Figs.~\ref{fg:a6}~--~\ref{fg:a8}). 
  
To explore the effects of lesser peaks on the derived
velocity for the two techniques, 
we compare XC and SEMLF results to the velocity of the brightest
component.  Fig.~\ref{fg:a6} shows the difference between the
velocity determined by each technique and the input model velocity of the
brighter component, as a function of $\Delta V$.  We compare the results
to the flux-weighted mean of the H$\alpha$ peaks (thin dotted line). 
The solid line shows the XC results; the dashed line 
shows the SEMLF results.  For small
$\Delta V$, the XC line is close to or above 
the SEMLF line --- the XC velocity is
generally closer to the velocity of the brighter component and SEMLF tracks the
flux-weighted mean.  At larger $\Delta V$, in models with 
larger component flux ratios, 
the lines cross --- SEMLF stops tracking the flux-weighted mean
and the SEMLF result is closer to
the velocity of the bright component.  
In the 4:3 flux ratio model,
SEMLF switches to fitting only the brighter peak at
$\Delta V \geq 280{\rm\ km\ s}^{-1}$ (not shown); in this model,
the greatest difference 
between the velocities from the two techniques is 44\,\%
of the component separation, when
$\Delta V = 240{\rm\ km\ s}^{-1} = 2.3$ times the FWHM of
H$\alpha$.
In summary, at low component separation, 
for all the models, SEMLF fits closer to the mean of the components,
while XC fits closer to the brightest peak.  At larger component 
separation, when the component flux ratio is large, this trend
reverses and SEMLF fits the brighter peak only.

Fig.~\ref{fg:a7} shows the errors computed by each technique
as a function of component velocity separation, for XC (solid
line) and SEMLF (dashed line --- unweighted; dotted line --- 
weighted by $\hat{\chi}^2$).  These figures illustrate
the necessity of weighting the formal SEMLF errors by $\hat{\chi}^2$.
The XC errors and the weighted SEMLF errors
increase dramatically as $\Delta V$ increases, indicating the presence of 
additional components, although the weighted SEMLF errors
begin to decrease again at very large $\Delta V$, for 2:1 or 4:1
flux ratios, when only the strongest peaks 
enter the fit.  In this case, the $\hat{\chi}^2$ remains large, as shown
in the next figure.

In Fig.~\ref{fg:a8}a and b, we illustrate the behavior of the XC
errors and of the $\hat{\chi}^2$ of the 
SEMLF Gaussian fits for flux ratios of 4:3 and 4:1; we plot the 
$r$ value on top, then $\omega$, then
$\hat{\chi}^2$.  In the 4:1 
flux ratio model $\omega$ increases until $\Delta V$ 
is so large that only one component contributes to the fit and
$\omega$ begins to decrease.  However, $r$
decreases with $\Delta V$ for each model at large
$\Delta V$, signaling the increasing
inadequacy of a single-component template as the velocity components
separate.  The decrease in $r$ accounts for the
increase in the errors; thus the errors increase monotonically as
$\Delta V$ increases.  Likewise, $\hat{\chi}^2$ increases 
rapidly as $\Delta V$
increases, slowing only at the very largest $\Delta V$, where the component
ratio is 4:1.

In these examples, XC and SEMLF compute velocities
which differ by up to 44\,\% of the component velocity separation.
XC errors increase
monotonically, and by a large factor, as the velocity component
separation ($\Delta V$) increases.  For larger component flux 
ratios, the unweighted SEMLF errors remain small and 
begin to decrease at large 
$\Delta V$, making them unsuitable for spectra with multiple velocity
components.  The $\hat{\chi}^2$ -- weighted SEMLF errors
behave much more like the XC errors, increasing by large 
factors as $\Delta V$ increases,
although they do not increase monotonically for large component flux
ratios with large $\Delta V$.  Note that because the $\hat{\chi}^2$ taken 
alone does increase monotonically, $\hat{\chi}^2$ is preferable to the
weighted error as an indicator of multiple velocity components when
using SEMLF.

The cross-correlation
technique flags the multiple-component case consistently, 
even as the secondary component becomes weak or widely separated 
in velocity from the primary component.
Cross-correlation errors are less model-dependent than formal
$\chi^2$-minimization errors in the sense
that they include mismatch between the model (the template)
and the spectrum. Thus in spite of a template-dependent systematic
bias in the errors, the cross-correlation errors roughly scale
properly; they even reflect velocity components that do not overlap
the strongest component in wavelength. 
The $\hat{\chi}^2$ of the SEMLF technique
also reflects the presence of secondary velocity components that
overlap the main component in wavelength.

\section{Two-Dimensional Model Rotation Curves}

Here, we model effects from multiple velocity
components that can be important in longslit galaxy spectroscopy.  We
compare the results of XC and SEMLF. 

\subsection{Non-Gaussian Profiles}

When neighboring discrete disk components of a galaxy are not
sufficiently resolved, they overlap spatially, 
resulting in spectra with multiple
velocity components.  Spectra with multiple velocity components also
occur when galaxies have kinematically distinct, cospatial components,
like circumnuclear disks.  We illustrate these two cases with
two-dimensional models of ``rotation curves''; we construct the models
using artificial one-dimensional emission-line templates, or sums of
templates, created with {\bf linespec} in RVSAO, joined with {\bf
mk1dspec} and {\bf mk2dspec} in NRAO.ARTDATA. 

\begin{figure}
\plottwo{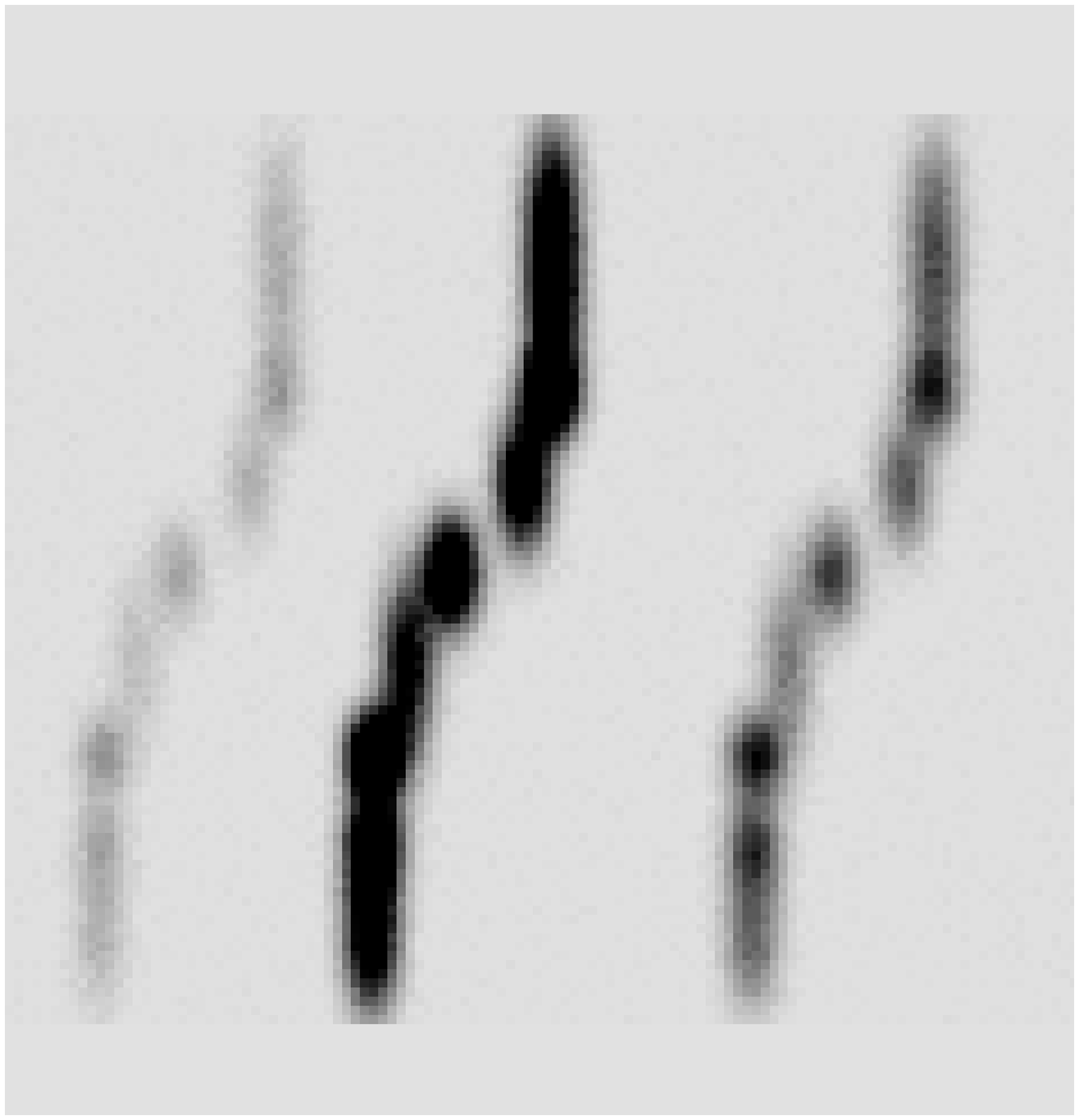}{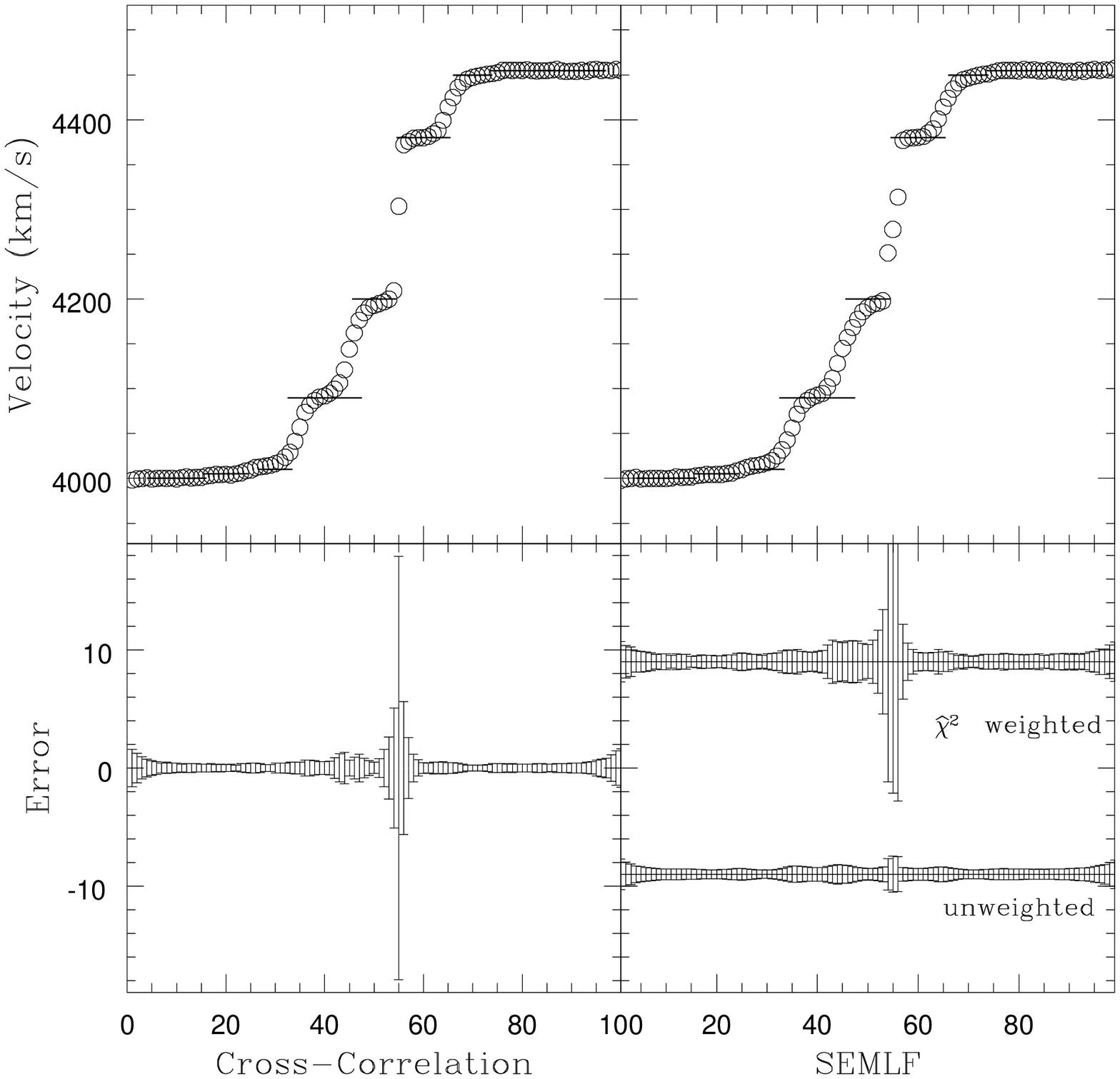}
\caption{``Overlapping segment'' spectral model: 
(a) greyscale image, showing the model H$\alpha$ and
Nitrogen lines; the dispersion axis is in the vertical 
direction and the spatial axis is horizontal and, (b) results and
errors for XC and SEMLF.  The top portion of (b) shows the rotation
curves superimposed on the model flux components (solid
horizontal lines), and the bottom portion shows the error bars
on an enlarged scale, including both weighted and unweighted
SEMLF errors.  The left sides of the rotation curves in 
(b) corresponds to the bottom of (a).}
\label{fg:mod0}
\end{figure}

Fig.~\ref{fg:mod0} is an image of a model spectrum which illustrates
spatially
overlapping components. Fig.~\ref{fg:mod0}a shows a greyscale plot of
the region around H$\alpha$ of the model longslit spectrum --- the
dispersion axis is horizontal, while the spatial axis is vertical and spans
100 pixels.  Our two-dimensional spectra consist of segments (the lumps
in the image) with Gaussian intensity variations in the spatial
direction.  The horizontal lines in the top portion of
Fig.~\ref{fg:mod0}b show each segment as a horizontal line
across its spatial FWHM, where the spatial direction 
is along the x-axis.  The actual ``emission'' extends 
beyond its FWHM.  The y-axis shows the velocity of each segment. 
The circles show the results from XC (left) and SEMLF (right), with
the errors on an expanded scale at the bottom of the figure.  We show both
unweighted SEMLF errors and SEMLF errors weighted by $\hat{\chi}^2$.

The velocity structure is Gaussian and does not vary along each
segment; line profiles are irregular (e.g. double-peaked) where the segments
overlap and thus more than one segment contributes to the spectrum. 
Widths in the dispersion direction are fixed by the basic spectrum,
Fig.~\ref{fg:a1}, used in the models.  As in the one dimensional case,
we add read noise (7.0~e$^{-}$) and Poisson noise to the spectra using
{\bf mknoise} in NRAO.ARTDATA. 
Although the model is a step function in velocity, the calculated
rotation curve varies smoothly because the segments at different velocities
overlap spatially.

The biggest differences between the XC and SEMLF curves
occur between segments, at pixel $\sim$55.  
The errors behave similarly for the two techniques. The errors for both
models increase significantly in the overlap regions; they are larger
where adjacent components are separated by larger velocities.  The weighted
SEMLF errors increase by a larger factor than the unweighted errors increase.
As expected, $\omega$ increases in the
regions of overlap, and $r$ decreases; $r$ also decreases when the
signal fades (e.g. at pixel values $> 95$). 

\begin{figure}
\plottwo{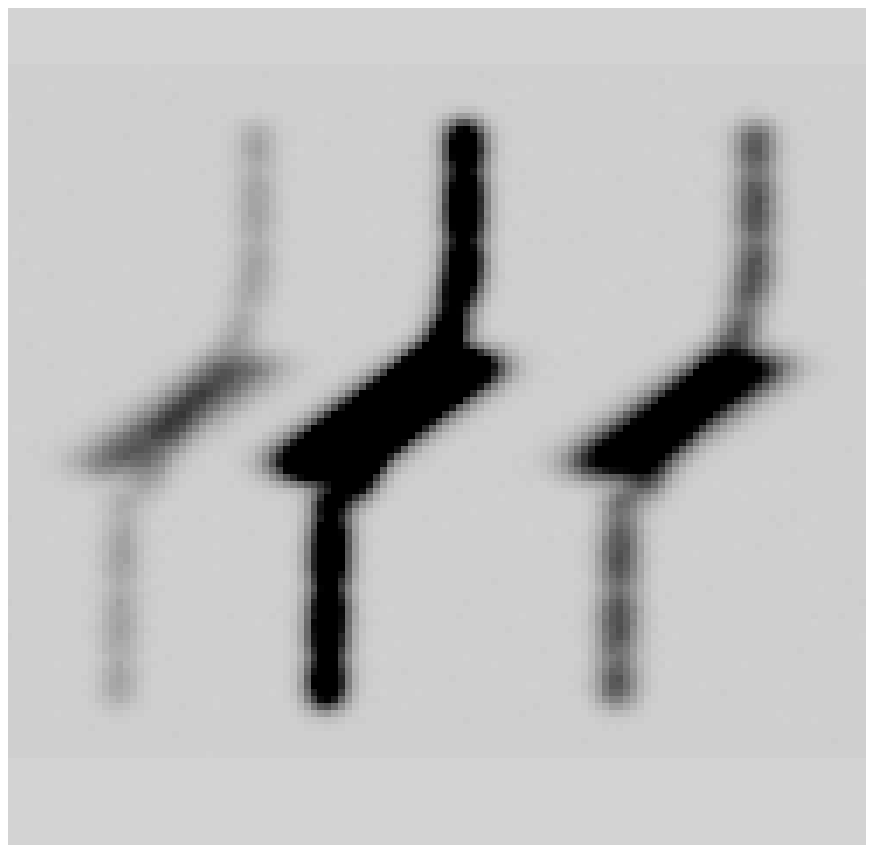}{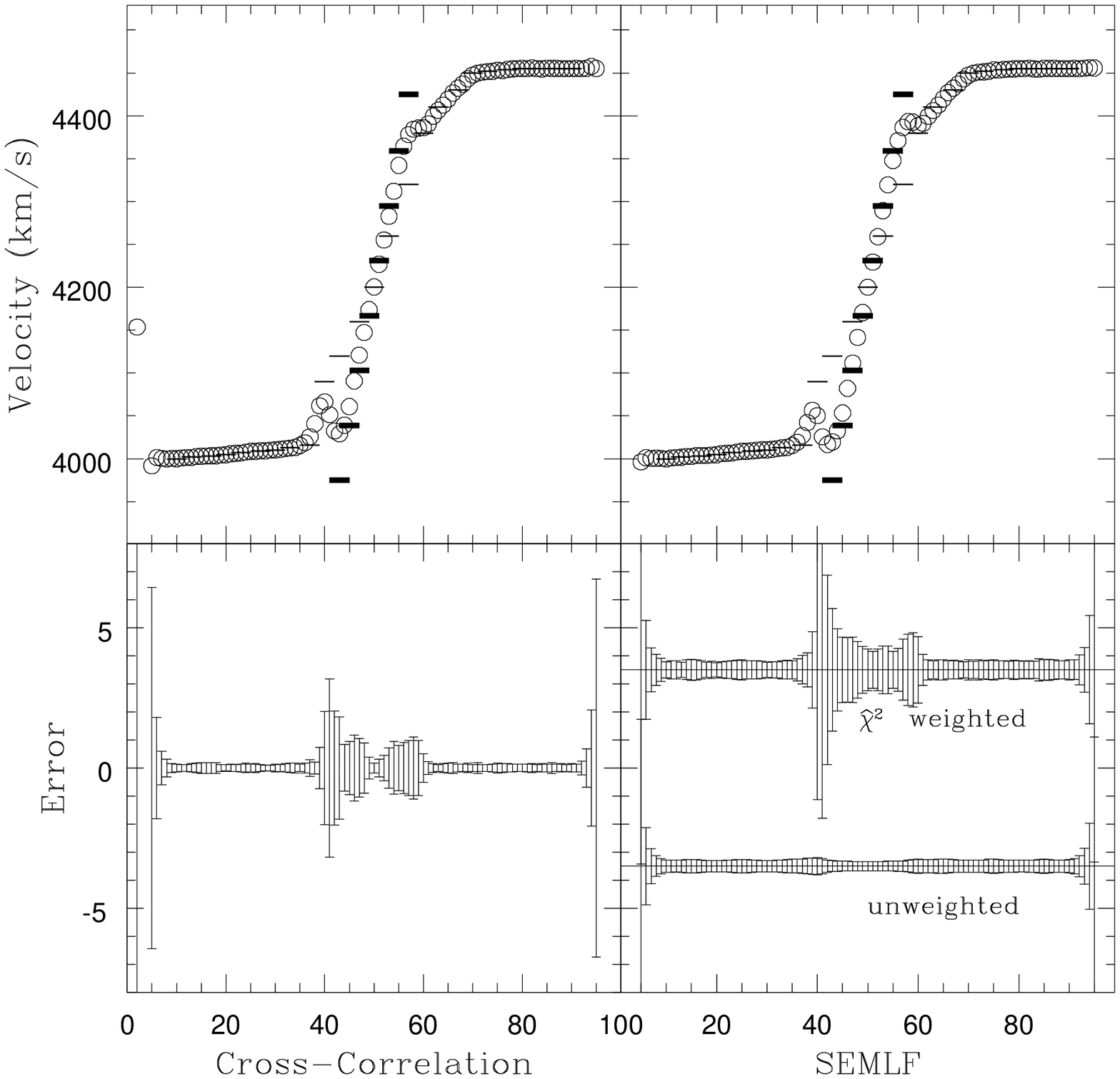}
\caption{``Inner disk'' spectral model (see Fig.~9 for a 
description of the figure, but note the different error
scale here).  Note that the XC and weighted
SEMLF errors increase in the center, reflecting the two
kinematic components, but the (formal) unweighted SEMLF errors
decrease in the center --- they fail to reflect the complex
velocity structure.}
\label{fg:mod2}
\end{figure}

The model in Fig.~\ref{fg:mod2}a resembles a 2-component galaxy
with an inner disk (see e.g. Rubin et al. 1997).  Fig.~\ref{fg:mod2}b
shows the positions and velocities of the segments; each segment has a
FWHM of 4 pixels in the spatial direction, although we plot only
points. The outer disk model (dashed line) approximately traces a
standard rotation curve.  The second model component (thick solid line)
represents an inner gas disk; it rotates as a solid body, is twice
as intense as the outer curve, and has emission lines twice as
broad. 

Fig.~\ref{fg:mod2}b shows that the techniques fit similar rotation 
curves, with similar errors, to the model.  In the
cross-correlation case, the $\frac{1}{1+r}$ contribution to the error
decreases in the center of the model due to larger signal, but 
{\it the cross-correlation errors increase} because $\omega$ increases,
signaling the presence of the second, distinct
velocity component (see Fig.~\ref{fg:mod2}b).

When the formal SEMLF errors are not weighted by $\hat{\chi}^2$, 
{\it the errors decrease in the center of the model} due to the larger
S/N ratio.  Only $\hat{\chi}^2$--weighted errors reflect the velocity
uncertainty due to the two components, because $\hat{\chi}^2$ increases 
as a single Gaussian becomes a poor fit.

\subsection{Nonstandard Line Ratios with Non-Gaussian Profiles}

\begin{figure}
\centerline{\epsfxsize=6in%
\epsffile{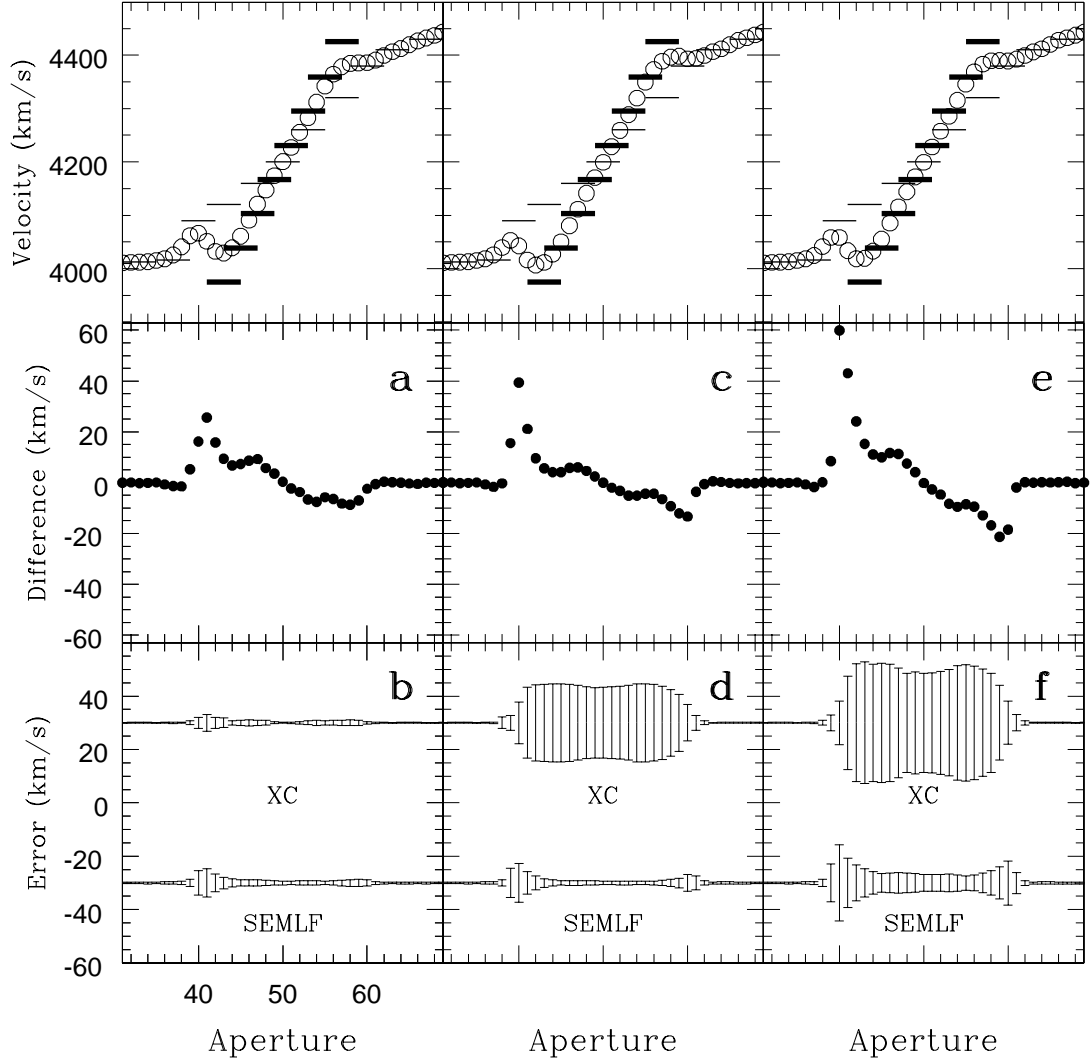}}
\caption{XC and SEMLF velocity differences and errors for the ``inner
disk'' model and models with varying line ratios for the inner disk
component.  The top figures are the XC rotation curves.
The middle figures are the SEMLF velocity
minus the XC velocity; the bottom figures are errors for each
technique.  (a) and (b) are the ``inner disk'' model from 
Fig.~\ref{fg:mod2}.  (c) and (d) are the same model, except the
[NII] line heights are increased by a factor of 10 for the wide
inner disk velocity component.  (e) and (f) also have [NII] heights
increased by a factor of 10, and no H$\alpha$ emission from the
inner disk component.}
\label{fg:a11}
\end{figure}

Nonstandard line ratios (e.g. in the nuclear regions of AGN) are another 
potential source of template or model mismatch.  
In real galaxies, non-thermal activity and multiple components
often arise together; we consider a combined model here.
We compare the XC and SEMLF
responses to nonstandard line ratios in the nuclear region using the
``inner disk'' model;  we plot results in Fig.~\ref{fg:a11}.
The left column of Fig.~\ref{fg:a11} shows the model of 
Fig.~\ref{fg:mod2} on top,
the XC velocity minus the SEMLF velocity
in the middle, and  the errors from the two
techniques on the bottom.  The middle column model has [NII] lines
with heights greater by a factor of ten in the inner disk component;
the outer disk and the [SII] and H$\alpha$ lines remain the same.  The
model we show in the right column has [NII] lines that are ten
times larger and no H$\alpha$ emission in the inner disk component.
Again, the outer disk and [SII] profiles remain the same --- thus,
there is a small amount of H$\alpha$ emission from the outer disk component.
The velocity structure of the models remains the same.

The changing line ratios influence the results of the techniques
somewhat, especially at the edge of the inner disk component. The 
peak difference between SEMLF and XC results occurs
at aperture 40 in Fig.~\ref{fg:a11}e, where the difference is
$60 {\rm\ km\ s}^{-1}$, corresponding to $52\,\%$ of the
velocity difference between the inner and outer disks at that aperture.
In XC, the contributions to the velocity from each emission line 
are effectively weighted by their heights in the template.  H$\alpha$ is
the strongest emission line in the HII region template we use.
Thus, XC weights the contribution of the faint outer disk 
much more heavily than SEMLF does, due to its small amount
of H$\alpha$ emission.
Peak heights in SEMLF may vary to accommodate
changing line ratios; thus the SEMLF errors increase only moderately.
The XC errors increase due to spectrum/template
mismatch.  When the mismatch is severe, the templates can be adjusted.
For example, when Balmer absorption
eliminates H$\alpha$, the line can be removed from the template.

Increased errors may result from many sources, 
including changing line ratios, 
lower S/N, or additional velocity components. All
three of these effects increase the $r$ error statistic in
XC.  $\omega$ reflects multiple velocity components that
are not spaced too widely.
To determine the cause of increased errors, it is usually necessary to
examine the line profiles in the region of interest.  However, 
the dip in the computed rotation curve at the transition between
the inner and outer disk (Fig.~\ref{fg:mod2}, aperture 43),
accompanied by the increased error, or especially by an
increased $\omega$, provides a
strong clue to the nature of the increase --- a second velocity
component.  In automated reduction with XC, one can select out 
rotation curves where there are many adjacent points with large 
errors as candidates for multiple-component systems.  With SEMLF,
the formal errors do not clearly reflect additional components;
they flag only regions of low S/N ratio.
The effects of spectrum/model mismatch are isolated by 
$\hat{\chi}^2$; regions with multiple velocity components
can be flagged as regions where $\hat{\chi}^2$ is significantly
greater than 1.

\section{Application of XC to Real Galaxies}

\begin{figure}
\plottwo{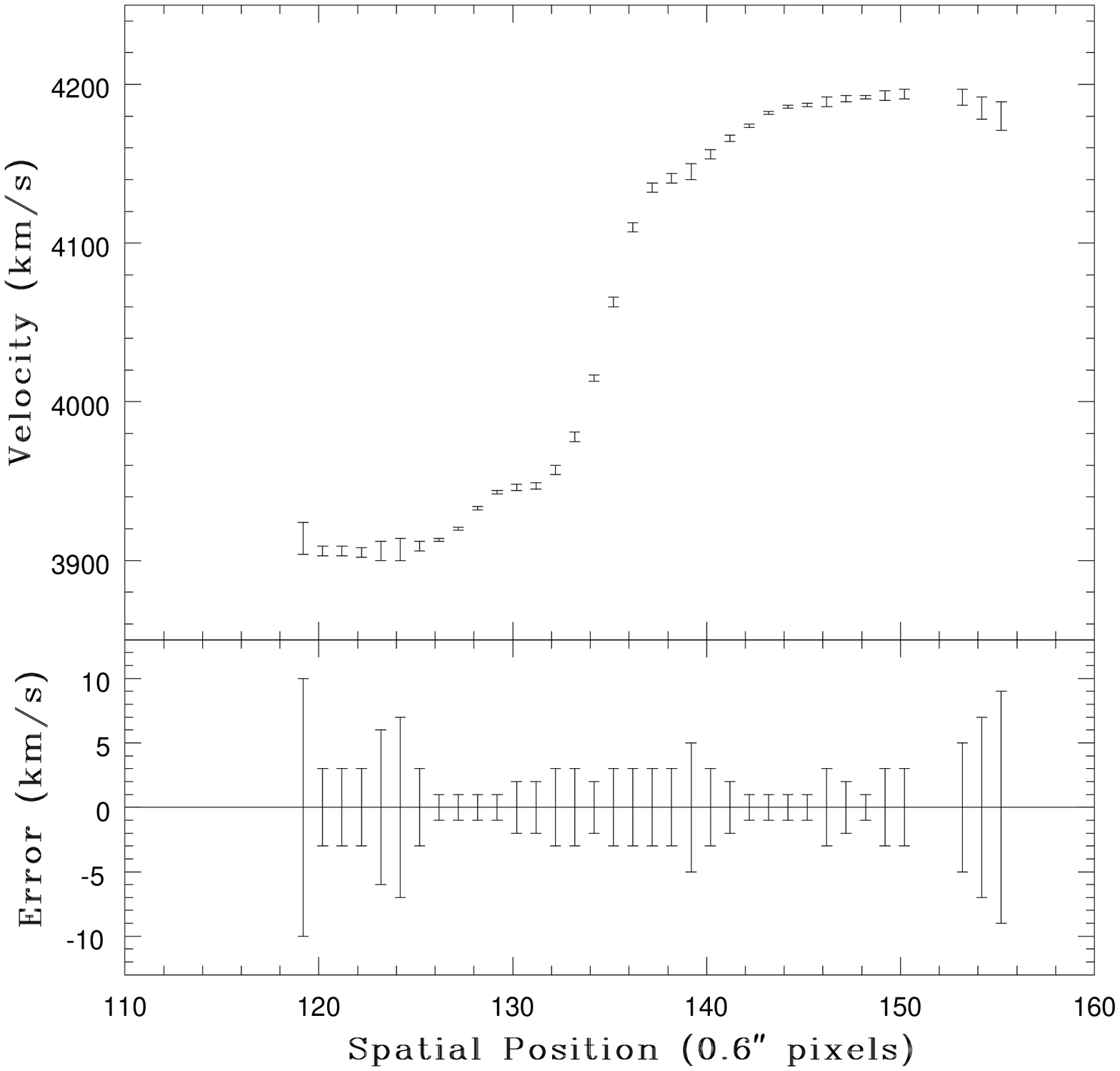}{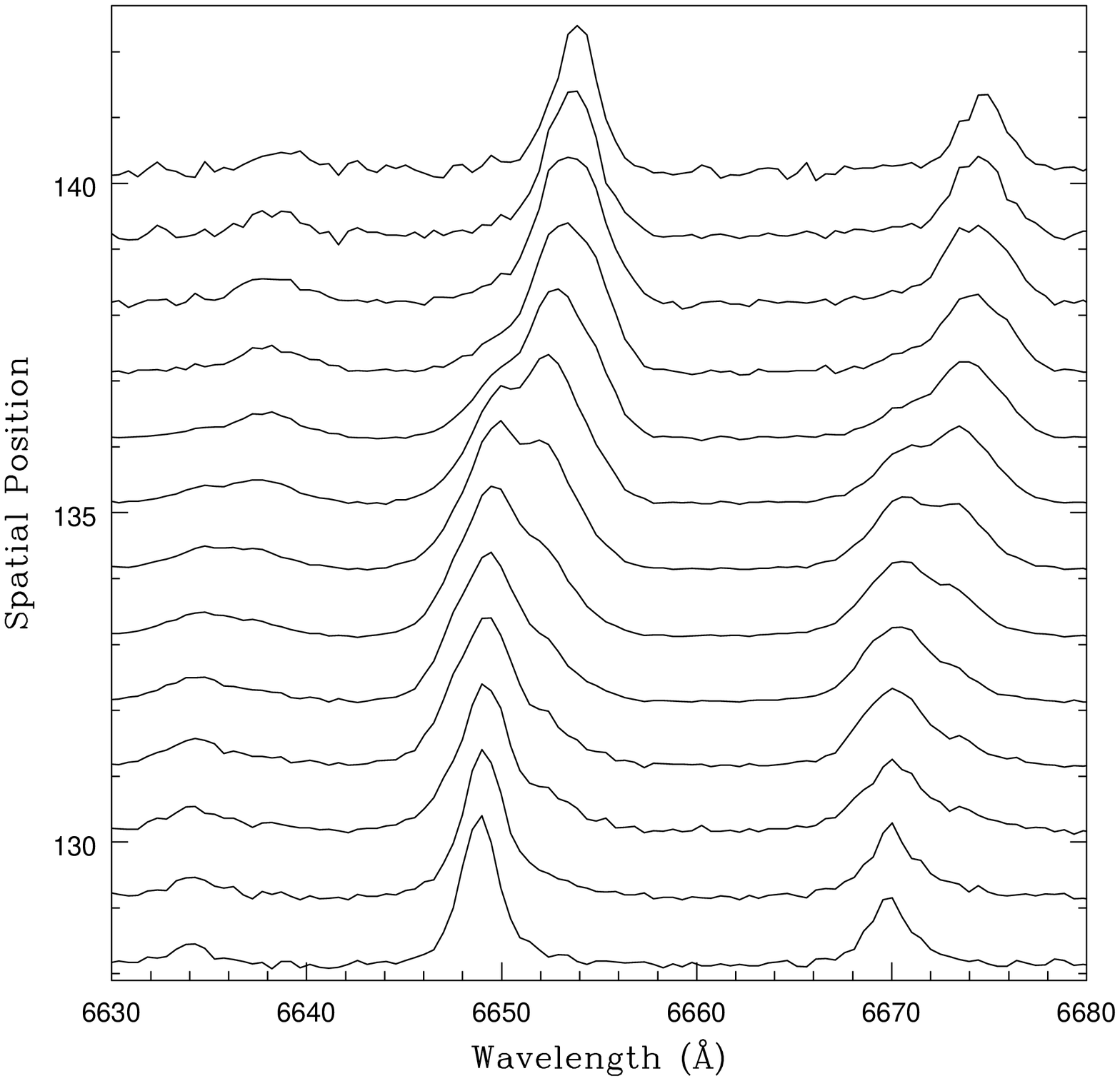}
\caption{The inner rotation curve of a real galaxy with 
a separate kinematic component in its center: (a) 
XC rotation curve and errors, and (b) Nitrogen and H$\alpha$
line profiles at spatial positions near the center 
of the galaxy.  Each profile is normalized for display
purposes --- the true emission lines
have more flux in the center than on the outskirts.
The labels on the y axis correspond
to the values on the x axis of (a).}
\label{fg:a12}
\end{figure}

Barton et al. (1999) use XC to reduce a large sample of
rotation curves of galaxies in pairs; 
Fig.~\ref{fg:a12}a shows an example.  Barton et al. (1999)
describe the reduction procedures in detail.  The curve shows the inner
part of CGCG 373-046, which has a separate kinematic
component in the center. The errors enlarge in the center
to reflect this component.  Fig.~\ref{fg:a12}b shows H$\alpha$ 
and [NII] line profiles at various slit positions; 
the line profiles are clearly doubled near
the center of the galaxy, where the errors enlarge. 

The models described in this paper 
explore only the simplest cases; they exclude the
effects of uncertainty in wavelength solution, night sky
contamination, cosmic rays, non-Gaussian line profiles other
than multiple-component profiles, continuum emission and extra
emission lines (that are not included in the template spectra).  
At minimum, the steps necessary for longslit redshift reduction are:
(1) bias subtraction and flat-fielding, (2) ``line-straightening,'' or
solving for the wavelength solution at each pixel in the spatial
direction, (3) cosmic-ray removal and sky subtraction, (4) a correction
for differential atmospheric refraction and (5) a redshift
determination at each point along the slit. 

Template selection or construction is also necessary, 
after step (4), when cross-correlation is used in step (5).  
The best template will reflect the spectrum of a typical 
single-velocity component.  Barton et al. (1999) 
build cross-correlation templates using the
observed relative line heights and widths. To measure line heights and
widths for the template, they run the task {\bf emsao} 
in the RVSAO package (Kurtz \& Mink
1998) which fits unconstrained Gaussian functions to 
the major emission lines with enough signal.  
They use each aperture in which {\bf emsao} finds all 5 emission
lines; they compute the median relative line heights and the median
absolute line widths for each run (4 -- 5 nights).  Barton et al.
use the resulting values as input parameters to create a template of smooth
Gaussian ``emission'' lines with {\bf linespec} in RVSAO.
They test the performance of separate templates
for each observing run, night, and galaxy and find that a run template
yields the smallest errors on average.  In practice, a single,
carefully constructed template used for all data from a 
particular instrumental setup should
suffice for most purposes (see Kurtz \& Mink 1998).

For the low S/N case, a small amount
of fine-tuning is necessary to extract complete rotation curves;
Barton et al. find that restricting the wavelength
range of allowable solutions to within $\sim$1000~km~s$^{-1}$
of the systemic velocity is easy to implement and yields
accurate curves in low S/N regions, although it may exclude
extreme cases of separate velocity components.   For example,
Morris et al. (1985) find a
component of NGC~7582 1300~km~s$^{-1}$ off the systemic
velocity.
.

\section{Conclusion}

We describe the use of cross-correlation to determine velocity fields
of nearby galaxies using optical emission lines observed with longslit
spectroscopy.  The method is easily automated,
makes simultaneous use of the strongest
emission lines, and is efficient for low signal-to-noise spectra.  
As we describe, the technique yields well-defined errors. 

We compare cross-correlation to a fundamentally different,
parametric technique, simultaneous Gaussian fitting of
emission lines. Velocities and errors computed by the two techniques agree
very well in the case of a single Gaussian velocity component.  

When line profiles are non-Gaussian (e.g. because more than one component
contributes at the same slit position), the results of the two techniques
differ.  In our examples,
the XC and SEMLF techniques give velocities which differ by up to
52\,\% of the component velocity separation.  
For standard HII region emission line ratios and
component separations up to 1.5 times
the FWHM of H$\alpha$, XC
fits closer to the brightest peak and SEMLF fits closer to the mean.
As the separation becomes larger, SEMLF also switches to the
brightest peak.

The formal SEMLF error and the error computed by XC differ 
significantly because the XC error 
consistently reflects mismatch of the spectrum and model (template), 
whereas formal SEMLF errors are model-dependent. 
However, when SEMLF errors are weighted by $\hat{\chi}^2$, 
SEMLF errors behave much more like XC errors.  
Only minor differences remain, as the
SEMLF errors do not increase monotonically in reflecting components
with increasing velocity separation from the primary velocity
component.  Thus, the $\hat{\chi}^2$-weighting procedure is 
empirically justified by comparison with XC.

For automated reduction of large data sets, 
multiple components and other non-Gaussian line structures
can be flagged for further inspection using the increase in
either the XC error (or the statistic $\omega$), or the
SEMLF $\hat{\chi}^2$.  However, a complete 
description of these multiple-component
cases {\it requires additional modeling} to explore the different
components.  

The choice of whether to use XC or SEMLF should be guided by the 
following considerations:
(1) XC is readily available as the IRAF routine {\bf xcsao}, in
the RVSAO package.  It is easily automated and requires
no initial redshift guess (although constraints on the allowed
solutions are useful for spectra with small S/N ratios),
(2) for two-component profiles, XC generally fits closer to the
brighter peak, whereas SEMLF shifts from fitting the flux-weighted
mean to fitting the brighter peak for larve $\Delta V$ and large
flux ratios,
(3) both XC errors and the SEMLF $\hat{\chi}^2$ statistic 
may be used to flag multiple components,
(4) the XC error increases monotonically with peak velocity
separation, as does the SEMLF $\hat{\chi}^2$,
but the SEMLF errors do not behave monotonically because
at large $\Delta V$, SEMLF switches to fitting only the
brightest peak,
(5) SEMLF errors are exact in the ideal Gaussian case, whereas the 
overall normalization of the XC errors must be calibrated for each
template if $\pm 20$\% errors are not accurate enough, 
(6) in XC the choice of template fixes the model
emission line ratios and linewidths, whereas these can vary (or be
fit), in SEMLF, 
and (7) SEMLF measures shape information for the line profiles 
in the single-Gaussian case.

\begin{acknowledgements}
We thank D. Fabricant, D. Mink and S. Tokarz for useful discussions
and assistance with the software.
E. B. and S. K. received support from Harvard Merit Fellowships.
E. B. received support from a National Science Foundation 
Graduate Research Fellowship, and S. K. received support from
a NASA Graduate Student Researchers Program Fellowship.  
This research was supported in part by the Smithsonian Institution.
\end{acknowledgements}

\clearpage

\clearpage

\end{document}